\documentclass[aps,pra,showpacs,preprint,superscriptaddress,floatfix,fleqn,byrevtex]{revtex4}
\usepackage[colorlinks=blue,breaklinks,bookmarks]{hyperref}
\usepackage{amsmath,amssymb}
\usepackage{graphics,color,epsfig}
\renewcommand{\Im}{\mathop{\rm Im}}
\renewcommand{\Re}{\mathop{\rm Re}}
\begin{document}
\title{Adiabatic passage and dissociation controlled by interference
of two laser-induced continuum structures }
\author{A.~K.~Popov}\email{popov@iph.krasn.ru}
\homepage{http://www.kirensky.ru/popov}
\affiliation{Institute of
Physics of the Russian Academy of Sciences, 660036 Krasnoyarsk,
Russia}
\affiliation{Office of the Chancellor/Departments of
Chemistry and Physics \& Astronomy, University of
Wisconsin-Stevens Point, Stevens Point, WI 54481, USA}
\author{V.~V.~Kimberg}\altaffiliation[Present address: ]{Theoretical Chemistry,
Department of Biotechnology, SCFAB Roslargstullbacken, 15 Royal
Institute of Technology (KTH), S-106 91 Stockholm, Sweden}
\email{viktor@theochem.kth.se} \affiliation{Institute of Physics
of the Russian Academy of Sciences, 660036 Krasnoyarsk, Russia}
\author{Thomas~ F.~ George}\email{tgeorge@uwsp.edu}
\homepage{http://www.uwsp.edu/admin/chancellor/tgeorge}
\affiliation{Office
of the Chancellor/Departments of Chemistry and Physics \&
Astronomy, University of Wisconsin-Stevens Point, Stevens Point,
WI 54481, USA}
\date{\today}
\begin{abstract}
We have developed a theory of three-pulse coherent control of
photochemical processes. It is based on adiabatic passage and
quantum coherence and interference attributed to the lower-lying
dissociation continuum and excited upper discrete states, which
are otherwise not connected to the ground state by one-photon
transitions. Novel opportunities offered by the proposed scheme
are demonstrated through extensive numerical simulations with the
aid of a model relevant to  typical experiments. The opportunities
for manipulating the distribution of the population among discrete
and continuous states with any necessary ratio by the end of the
pulses  are demonstrated.
\end{abstract}
 \maketitle
\section{Introduction}
Laser control of chemical reactions and other dissipation
processes like photodissociation of molecules and photoionization
of atoms has recently gained increasing interest in the laser and
chemical communities \cite{TR1,BrSh,TR2,Br1,Sh1,
Br2,R1,Br3,Br4,Br5,R4,Gor1,Sh2,Sh3, Mcool,R5,R6,R9,R8}. In many
cases, such control is based on the interference of different
quantum pathways.  While the main stream of theoretical and
experimental work on coherent quantum control concerns
laser-induced transitions between discrete states, coherence and
interference effects attributed to dissociation and ionization
energy continua are of great importance, because of various
applications in atomic and molecular physics and photochemistry.
The properties of the continuum of quantum states such as observed
in the ionization of an atom or  dissociation of a  molecule have
attracted interest since the first formulation of quantum theory.
Despite the fact that the shape of an autoionizing resonance is
stipulated by quantum interference \cite{Fan61,FanCo}, for a long
time the continuum was regarded as an incoherent dissipation
medium. It was until the work \cite{Fene61,GP1,GP2,GP3,GP4} which
established that the mixing of discrete and continuous states in a
strong, driving electromagnetic field led to similar constructive
and destructive interference processes. Thus the opportunity was
shown to produce an autoionizing-like resonance embedded in an
otherwise unstructured continuum, which displays itself in the
variety of photo- and nonlinear-optical processes similar to real
autoionizing states \cite{GP2,GP3,GP4,GP5,GP6,Gel}. The attractive
advantage of this option is that such laser-induced continuum
structures (LICS) can be produced, at least in principal, in any
necessary area of a continuum, and their strength is controlled
with the dressing laser. Since the first successful experiments
\cite{GP7,GP8,GP9,GP10}, which confirmed the principal theoretical
predictions, a great progress has been achieved in developing a
profound understanding of continuum coherence and related
laser-induced processes
\cite{GP11,Zol,Rza81,Pav82,GP12,Rza83,GP13,Kn84,KnLICS,Lam87,Hut,LamCom,Kn,Char91,
Cav91,Fau93,Char93,Fau94,CavJ95,Cav95,bull,Car,A2,Cav97,Kimb,Fau99}.

The appearance and consequences of coherence and interference
processes are different in the continuous wave regime, where the
relaxation processes play a crucial role, and in the pulsed regime
of rapid adiabatic passage
\cite{D1,D2,U1,U2,1D,1DD,D4,D3,D5,D8,D9,R2,R3,102,Band,Ric00,Br6,R7,117,YatOC02}.
Extensive studies of population transfer between two discrete
states via the upper-lying continuum in Raman-like
$\Lambda$-schemes have revealed that, despite the detrimental
effects, almost complete population transfer and corresponding
photoionization suppression is possible in such schemes, provided
that the driving laser pulses are properly ordered
\cite{Lam94,Kn97,U5,Kn98,A5,Half,Yats,U9,105}. It was also found
that coherence processes play an important role in two-photon
dissociation mediated by an intermediate resonance with a discrete
level allowing for the control of photodissociation and
photoassociation \cite{Sh94,Sh95,Sh96,ShV96,Sh97,Berg}. Population
transfer and dissociation play an important role  as the
accompanying processes in the three-pulse four-wave mixing
technique, which is used for spectroscopy and coherent quantum
control in chemistry \cite{L22,L23,L24,L25,L28,L29,L32}.

Based on the outlined achievements, this paper further develops
the theory of coherent quantum control, which employs LICS and
rapid adiabatic passage. We propose and investigate a more complex
three-pulse scheme (Fig.~\ref{lev}), which allows additional
flexibility and means for such control. Unlike the so far
investigated schemes, we report on novel opportunities for
manipulating dissociation and population redistributions between
the \textit {upper} bound states $n$ and $f$ through the interplay
of two LICS attributed to \textit {lower-lying} energy continuum.
Only one discrete state here is connected to the ground state by
one-photon transitions. All radiations are assumed to be Gaussian
pulses with variable sequence and duration, strong enough to drive
molecular transitions. Corresponding equations for slowly-varying
probability amplitude are derived, and extensive numerical
simulations are performed aimed at a demonstration of the outlined
opportunities under typical experimental conditions.

This paper is organized as follows. The set of equations for
slowly-varying probability amplitudes pertinent to the problem
under consideration are derived in Sec. \ref{be}. By turning off
the first or third laser, our scheme can be easily reduce to those
investigated in \cite{Half,Berg}, where excellent agreement
between theory and experiment was reported. Corresponding
analytical solutions are found for further analysis of
time-dependent destructive and constructive interference and for
testing the theory by comparing with the known results. They are
also useful for determining the dependence of such processes on
the complex effective Fano parameters, which represent the
properties of the specific continua. The results of various
numerical experiments are presented in in Sec. \ref{ns}.
Suppression of photodissociation through continuum coherence and
overlap of two LICS is simulated in Sec. \ref{2l}. The features of
photodissociation controlled by two ordered pulses are simulated
in Sec. \ref{pd}. We investigate the possibility of manipulating
the dissociation yield and populations of excited states, driven
by three pulses with variable sequence and duration.  Our goal is
to demonstrate the feasibility of achieving almost any necessary
ratio between the branching yields by the end of the pulses due to
the interference of quantum pathways through a variety of
continuum states.  This is done with the aid of numerical
experiments in Sec. \ref{tp}. We show how the pulse parameters
must be adjusted in order to enable such opportunities.

\section{Basic Equations}\label{be}
\begin{figure}[!h]
\includegraphics [width=0.35\textwidth] {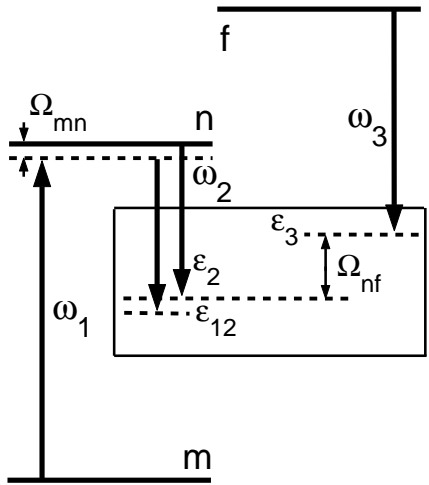}
\caption{\label{lev} Energy levels and driving fields.}
\end{figure}

The proposed coupling scheme is illustrated in the energy level
diagram depicted in Fig.~\ref{lev}.
Radiation at frequency $\omega_1$ couples the bound-bound
transition $m-n$, and radiations at  $\omega_2$ and $\omega_3$
couple the bound states $n$ and $f$ with  states of the
dissociation continuum $\varepsilon$, as shown in the picture.
Initially, only the lowest level is assumed populated. At
$\omega_1$ tuned near the one-photon resonance, stepwise and
two-photon transitions interfere. The two-photon transition rate
between the levels $n$ and $f$ depends on the detuning
$\Omega_{nf}$. Actually, all detunings are intensity-dependent,
which determine the complex dynamics of the interference between
various stepwise and multiphoton processes. We assume that each
laser drives only one transition and accounts  only for resonant
coupling depicted in Fig.~\ref{lev}. Therefore, we ignore other
off-resonant and incoherent processes caused by the driving field.
We also assume a single continuum, i.e., no degeneracy, in order
to avoid unnecessary complexity related to the processes, which
are not the subject of this paper. Basically, the set of equations
depends on whether the system is open or closed. In the case of a
closed model, the lowest level $m$ is the ground state, and the
sum of the level populations is always equal to 1. Alternatively,
in an open scheme, level $m$ is an excited state that relaxes, and
the sum of the populations depletes with time. For the purposes of
discussion, we will consider the second case, although for pulses
considerably shorter than all the relaxation times, the difference
between the closed and open schemes disappears. In the
rotating-wave approximation, the equations for slowly-varying
probability amplitudes in the case of an open energy-level system
are written as
\begin{align}
&i\dot a_{\varepsilon}=G_{\varepsilon n}\exp(i\Omega_{\varepsilon
n}t)a_n +
G_{\varepsilon f}\exp(i\Omega_{\varepsilon f}t)a_f\nonumber\\
&i\dot a_n+i\gamma_n a_n =G_{nm}\exp(i\Omega_{nm}t)a_{m}+\int
d\varepsilon
G_{n\varepsilon}\exp(i\Omega_{n\varepsilon}t)a_{\varepsilon},\nonumber\\
&i\dot a_f+i\gamma_f a_f =\int d\varepsilon
G_{f\varepsilon}\exp(i\Omega_{f\varepsilon}t)a_{\varepsilon},\nonumber\\
&i\dot a_m+i\gamma_m a_m =G_{mn}\exp(i\Omega_{mn}t)a_{n}.
\label{pu1}
\end{align}
Here $G_{\varepsilon n}=E_2d_{\varepsilon n}/2\hbar$,
$G_{\varepsilon f}=E_3d_{\varepsilon f}/2\hbar$, and
$G_{mn}=E_1d_{mn}/2\hbar$ are coupling Rabi frequencies;
$\Omega_{mn}=\omega_1-\omega_{mn}$,
$\Omega_{n\varepsilon}=\omega_{2}-\omega_{n\varepsilon}$ and
$\Omega_{f\varepsilon}=\omega_{3}-\omega_{f\varepsilon}$ are
frequency resonance detunings; $\omega_{i\varepsilon}=(E_i
-\varepsilon)/\hbar$; $E_i$ and $\varepsilon$ are the energies of
the corresponding discrete and continuum states; and $2\gamma_i$
is the decay  rate of level $i$. Following \cite{Gel}, we
introduce the values
\begin{align}
a_{\varepsilon}(t)&=\overline
a_{\varepsilon}\exp(i\Omega_{\varepsilon\varepsilon_0}t),\quad
a_{n}(t)=\overline a_{n}\exp(i\Omega_{n\varepsilon_0}t),\nonumber\\
a_{f}(t)&=\overline a_{f}\exp(i\Omega_{f\varepsilon_0}t),\quad
a_{m}(t)=\overline a_{m}\exp(i\Omega_{m\varepsilon_0}t),
\label{sabs}
\end{align}
where $\overline a_{\varepsilon}$ is the amplitude, which slowly
varies over the continuum  in the vicinity of the energy
$\varepsilon_0=E_f-\hbar\omega_3$,
\begin{align}
\Omega_{f\varepsilon_0}=\omega_{f\varepsilon_0}-\omega_3=0,\quad
\Omega_{n\varepsilon_0}=\omega_{n\varepsilon_0}-\omega_2=
\omega_{f\varepsilon_0}-\omega_{nf}-\omega_2=\Omega_{nf},\nonumber\\
\Omega_{m\varepsilon_0}=\omega_{m\varepsilon_0}-\omega_1+\omega_2=
\omega_{mn}-\omega_{n\varepsilon_0}-\omega_{1}+\omega_2=\Omega_{mn}-\Omega_{nf}.
\end{align}
The energy $\varepsilon_0$ is chosen in the vicinity of $E_n-\hbar
\omega_2$, if $E_3=0$.  We assume that the coupling parameters
(energy density of the transition oscillator strengths
$f_{i\varepsilon}$) vary around $\varepsilon_0$ substantially only
over the continuum energy intervals $\widetilde\varepsilon$, which
are much greater than the maximum of the characteristic widths of
the discrete states, including power-broadening $\hbar |V_{jk}|$
and pulse spectral width $\tau^{-1}$:
\begin{align}
\widetilde\varepsilon\gg\hbar\gamma_i, \hbar/\tau,\quad   \hbar
|V_{jk}|{\partial^2 f}/{\partial \varepsilon^2_{\varepsilon=
\varepsilon_0}}\ll {\partial f}/{\partial
\varepsilon_{\varepsilon=\varepsilon_0}}.\label{f}
\end{align}
Such requirements are fulfilled practically for all realistic
atomic and molecular continua in an energy range that is well
above the ionization or dissociation threshold. Since interference
processes appear at $|\Omega_{mn}|$, $|\Omega_{nf}|$ and
$|\Omega_{mn}-\Omega_{nf}|$ less than or on the order of the same
value, this enables us to separate the slow and fast variables.
With the aid of (\ref{sabs}), we obtain from (\ref{pu1})
\begin{align}
&i\dot{\overline
a_{\varepsilon}}-\Omega_{\varepsilon\varepsilon_0}{\overline
a_{\varepsilon}}=G_{\varepsilon n}{\overline a_{n}}+
G_{\varepsilon f}{\overline a_{f}},\label{pu2}\\
&i\dot{\overline a_{n}}-{\overline a_n} (\Omega_{nf}-i\gamma_{n})=
G_{nm}{\overline a_{m}}+\int d\varepsilon
G_{n\varepsilon}{\overline a_{\varepsilon}},\label{pu3n}\\
&i\dot{\overline a_{f}}+i{\overline a_f}\gamma_{f}= \int
d\varepsilon G_{f\varepsilon} {\overline a_{\varepsilon}},\label{pu3f}\\
&i\dot{\overline a_{m}}-{\overline a_m}
(\Omega_{mn}-\Omega_{nf}-i\gamma_{m}) = G_{mn}{\overline
a_{n}}.\label{pu3m}
\end{align}
The solution of  (\ref{pu2}) is
\begin{align}
{\overline a_\varepsilon}=e^{-i
\Omega_{\varepsilon\varepsilon_0}t}[C_1+\int(G_{\varepsilon
n}{\overline a_n}+G_{\varepsilon f}{\overline a_f})e^{i
\Omega_{\varepsilon\varepsilon_0}t}dt].\nonumber
\end{align}
Taking  slowly-varying values $G_{\varepsilon i}{\overline a_i}$
(as compared with the oscillating exponents) out of the integral,
we obtain
\begin{align}
{\overline a_\varepsilon}=e^{-i
\Omega_{\varepsilon\varepsilon_0}t}[C_1+(G_{\varepsilon
n}{\overline a_n}+G_{\varepsilon f}{\overline a_f})\int e^{i
\Omega_{\varepsilon\varepsilon_0}t}dt]=e^{-i
\Omega_{\varepsilon\varepsilon_0}t}C_1-(G_{\varepsilon
n}{\overline a_n}+G_{\varepsilon f}{\overline
a_f})/\Omega_{\varepsilon\varepsilon_0}. \nonumber
\end{align}
It follows from the initial conditions that $C_1=0$, and therefore
the solution of (\ref{pu2}) is
\begin{align}
{\overline a_{\varepsilon}}=-(G_{\varepsilon n}{\overline a_{n}}+
G_{\varepsilon f}{\overline a_{f}})/\Omega_{\varepsilon \varepsilon_0}.
\end{align}
This can be derived also directly from (\ref{pu2}) because the
inequality $|\dot{\overline
a_{\varepsilon}}|\ll|\Omega_{\varepsilon\varepsilon_0}{\overline
a_{\varepsilon}}|$ is correct for the major continuum interval.
Further, with the aid of the $\zeta $-function,
\begin{align}
[i(\omega_{k}-\omega _{j\varepsilon})]^{-1}= \pi \delta
(\omega_{k}-\omega _{j\varepsilon})- i{\cal P}(\omega_{k}-\omega
_{j\varepsilon})^{-1}; \quad k=1,2,3;\quad j=m,n,f, \label{15}
\end{align}
where ${\cal P}$ stands for the principal value of an integral,
Eqs. (\ref{pu3n})-(\ref{pu3m}) can be presented in the form
\begin{align}
d{\overline a_m}/dt&=-iG_{mn}{\overline a_n}-
(\gamma_m+i\Omega_{m\varepsilon_0}){\overline a_m},
\nonumber\\
d{\overline a_n}/dt&=-iG_{mn}^*{\overline
a_m}-\gamma_{nf}(1+iq_{nf}){\overline a_f}-[\gamma_n+\gamma_{nn}+
i(\Omega_{n\varepsilon_0}+\delta_{nn})]{\overline a_n},\label{pu4}\\
d{\overline a_f}/dt&=-\gamma_{fn}(1+iq_{fn}){\overline
a_n}-(\gamma_f+\gamma_{ff}+ i\delta_{ff}){\overline a_f}.\nonumber
\end{align}
where
\begin{align}
&\gamma_{nn}=\pi\hbar G_{n\varepsilon_{0} }G_{\varepsilon_{0} n}
+\Re\left(G_{nk}G_{kn}/p_{kf}\right),\nonumber\\
&\delta_{nn}=\hbar{\cal P}\int d\varepsilon \cdot G_{n\varepsilon
}G_{\varepsilon
n}/(\varepsilon_{0}-\varepsilon)+\Im\left(G_{nk}G_{kn}/p_{kf}\right);\nonumber\\
&\gamma_{ff}=\pi \hbar G_{f\varepsilon_{0} }G_{\varepsilon_{0} f}
+\Re\left(G_{fk}G_{kf}/p_{kf}\right),\nonumber\\
&\delta_{ff}=\hbar{\cal P}\int d\varepsilon \cdot G_{f\varepsilon
}G_{\varepsilon
f}/(\varepsilon_{0}-\varepsilon)+\Im\left(G_{fk}G_{kf}/p_{kf}\right);\nonumber\\
&\gamma_{nf}=\pi \hbar G_{n\varepsilon_{0} }G_{\varepsilon_{0} f}
+\Re\left(G_{nk}G_{kf}/p_{kf}\right),\\
&\delta_{nf}=\hbar{\cal P}\int d\varepsilon \cdot G_{n\varepsilon
}G_{\varepsilon
f}/(\varepsilon_{0}-\varepsilon)+\Im\left(G_{nk}G_{kf}/p_{kf}\right);\nonumber\\
&\gamma_{fn}=\pi \hbar G_{f\varepsilon_{0} }G_{\varepsilon_{0} n}
+\Re\left(G_{fk}G_{kn}/p_{kf}\right),\nonumber\\
&\delta_{fn}=\hbar{\cal P}\int d\varepsilon \cdot G_{f\varepsilon
}G_{\varepsilon
n}/(\varepsilon_{0}-\varepsilon)+\Im\left(G_{fk}G_{kn}/p_{kf}\right);\nonumber\\
&p_{kf}=\Gamma_{kf}+i(\omega_{kf}-\omega_3),\quad
q_{ij}=\delta_{ij}/\gamma_{ij}, \quad i,j=n,f.
\end{align}
Besides the continuum states, the contribution of other
non-resonant levels $k$ is taken into account,  and a sum over the
repeating $k$ index is assumed. Contributions from these levels
may occur comparable to those of the continuum states. As seen
from the equations (\ref{pu4}), the values $\gamma_{ij}$ and
$\delta_{ij}$ describe light-induced broadening and shifts of
discrete resonances stipulated by the induced transitions between
them through the continuum. The magnitude of the shift and its
sign are determined by the overall counterbalance of the continuum
states below and above $\varepsilon_0$. The parameters
$q_{ij}=\delta_{ij}/\gamma_{ij}$ are analogous to the Fano
parameters for autoionizing states. They characterize the relative
integrated contribution of all off-resonant quantum states
compared to the resonant ones. Within the validity of (\ref{f}),
their dependence  on the field intensities can be neglected. The
ratio of the real and imaginary parts (relative phase of the
corresponding induced atomic oscillations) plays a crucial role in
whether the interference is constructive or destructive. This
depends on the effective Fano parameters $q_{ij}$, which are
determined by the distribution of the oscillator strengths over
the continuum and discrete states and on the positions of the
resonant continuum states, and can be controlled through
multiphoton detunings.

Further, we will use dimensionless variables, scaled to the pulse
$|E_1|^2$ half-duration $\tau$ at the $1/e$ level: $T=t/\tau$,\,
$\gamma_i\tau=\eta_i$,\, $g_{mn}=G_{mn}\tau$,\,
$g_{nf}=\gamma_{nf}\tau$,\, $g_{nn}=\gamma_{nn}\tau$,\,
$g_{ff}=\gamma_{ff}\tau$,\, $\Delta_{mn}=\Omega_{mn}\tau$,\,
$\Delta_{nf}=\Omega_{nf}\tau$. Then the equations take the form
\begin{align}
d\overline a_m/d T&=-ig_{mn}\overline a_n-[\eta_m+i(\Delta_{mn}-
\Delta_{nf})]\overline a_m,\nonumber\\
d\overline a_n/d T&=-ig_{mn}^*\overline
a_m-g_{nf}(1+iq_{nf})\overline a_f-[\eta_n+g_{nn}+
i(\Delta_{nf}+q_{nn}g_{nn})]\overline a_n,\label{a}\\
d\overline a_f/d\tau'&=-g_{fn}(1+iq_{fn})\overline
a_n-(\eta_f+g_{ff}+ iq_{ff}g_{ff})\overline a_f.\nonumber
\end{align}
With the aid of these equations, the dissociation probability is
calculated as
\begin{align}
W=2\int d T\left\{(g_{nn}|\overline a_n|^2+g_{ff}|\overline
a_f|^2+2\Re\left[g_{nf}\overline a_n\overline
a_f^*\exp(i\Delta_{nf}T)\right]\right\}.\label{di}
\end{align}
The first two terms, proportional to the squared moduli of the
probability amplitudes, describe the multistep dissociation
associated with the populations of levels $n$ and $f$. The third
term describes quantum control through the interference of
coherent quantum pathways. As seen from Eqs. (\ref{pu4}) and
(\ref{a}), the magnitude and sign of the interference term in
(\ref{di}) strongly depends on the phases of the probability
amplitudes  and, consequently, on the parameter $q_{fn}$.


\section{Numerical simulation of two- and three-pulse coherent
control}\label{ns} Many experiments on the coherent control of
branching chemical reactions have been carried out with  sodium
dimers Na$_2$. In this case, level $m$ of our model (Fig.
\ref{lev}) can be attributed to the state $X^1\Sigma_g^+
(v=28,J=10)$,  and levels $n$ and $f$  to the states
$A^1\Sigma_u^+ (v=37,J=11)$ and $A^1\Sigma_u^+ (v=47,J=11)$, which
are coupled by strong transition dipoles with the dissociation
continuum Na(3s)+Na(3s) \cite{Sh1}. Alternatively, vibrational
states of the electronic excitation $B^1\Pi_u (v,J=11)$ can be
chosen as level $f$. The characteristic relaxation rates of these
states are: $\gamma_m=2\cdot 10^7 \,c^{-1}$, $\gamma_n=
\gamma_f=1.2\cdot 10^8\, c^{-1} $. We assume all pulses to be
Gaussian,
\begin{align}
g_{mn}&=g_{mn}^0\exp(- T^2/2),\nonumber\\
\gamma_{nn}&=\gamma_{nn}^0\exp\left[-(T -
\Delta_2)^2/d_2^2\right],\label{g}\\
\gamma_{ff}&=\gamma_{ff}^0\exp\left[-(T -
\Delta_3)^2/d_3^2\right],\nonumber
\end{align}
where $\Delta_2$ and $\Delta_3$ are the delays of the pulses $E_2$
and $E_3$ with respect to the pulse $E_1$, $d_2=\tau_2/\tau$ and
$d_3=\tau_3/\tau$ are the durations of pulses  $E_2$ and $E_3$
($\tau_2$ and $\tau_3$) scaled to the duration of the first pulse
$\tau$. These values must be selected within a time domain range
shorter than $\tau=10^{-9}s$ in order to avoid relaxation over the
whole period of excitation.  As seen from Eq. (\ref{di}), the
overall analysis of the control reduces to an analysis of the
interference and multistep terms.
\begin{figure}[!h]
\begin{center}
{\bf(a)}\hspace{80mm} {\bf(b)}\\
\includegraphics[width=0.3\textwidth]{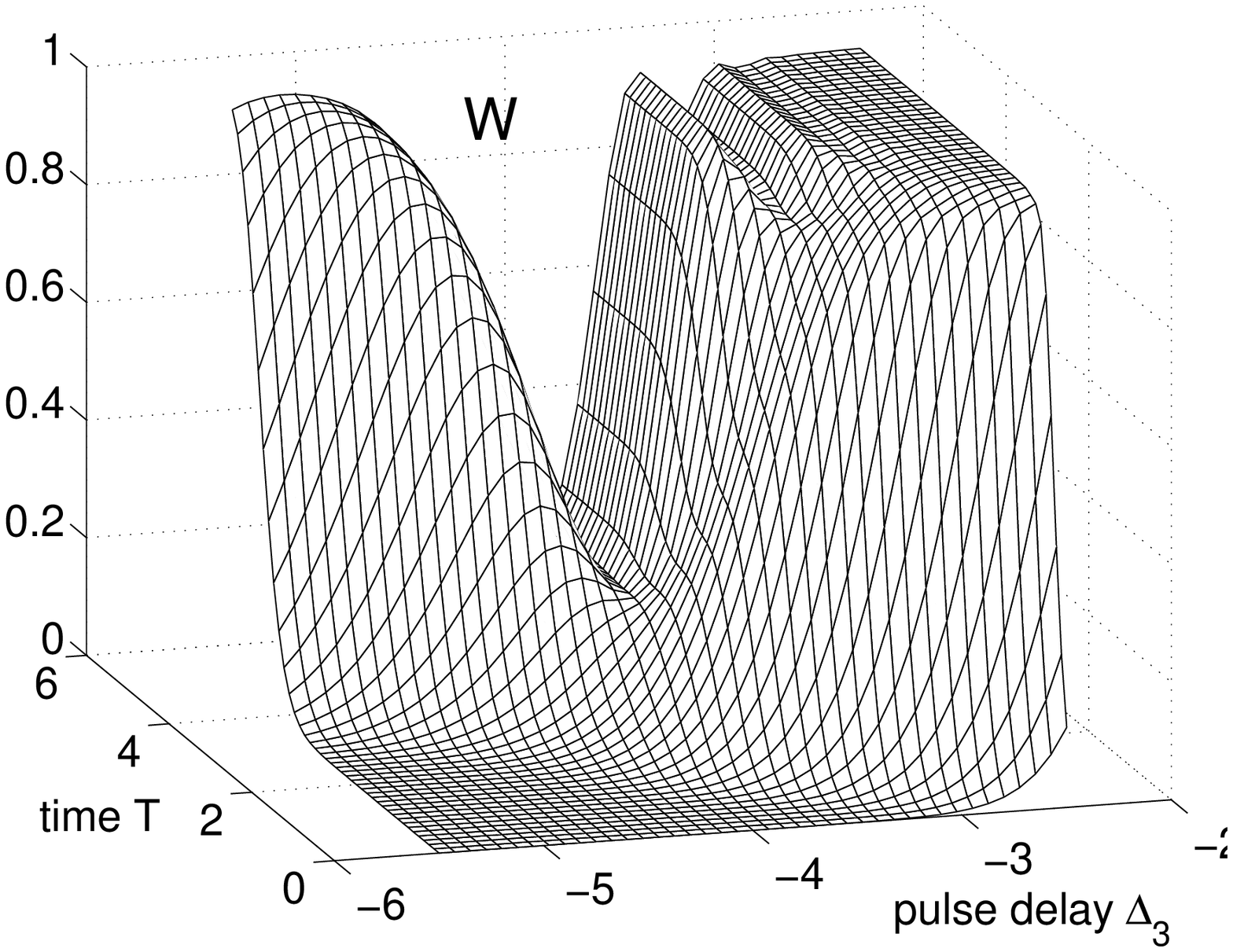}\hspace{5mm}
\includegraphics[width=0.3\textwidth]{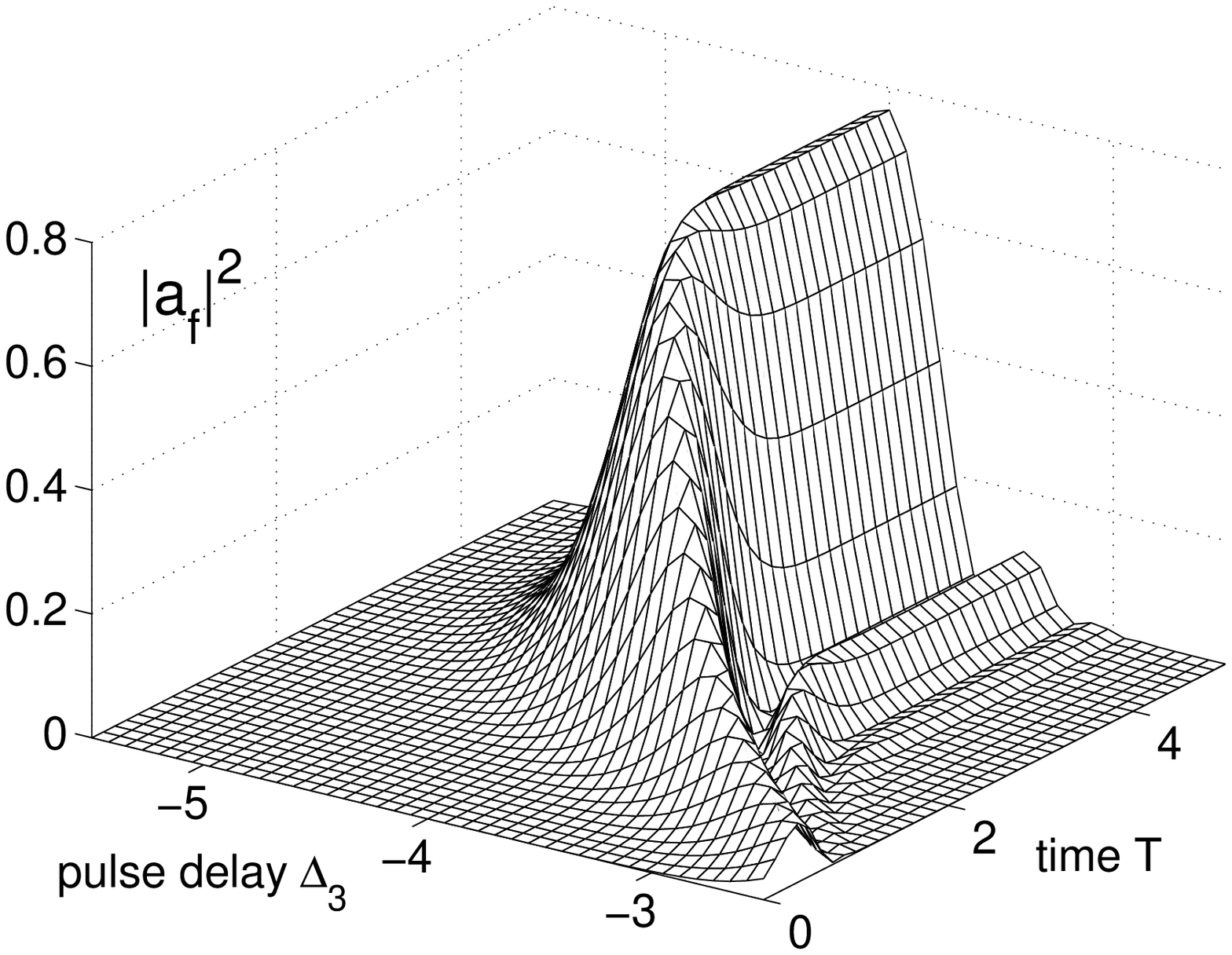}
\end{center}
 \caption{\label{p01} Dissociation (a) and population
transfer from level $n$ to level $f$ (b) vs time and delay of the
pulse $E_2$ relative to $E_3$ ($\Delta_2=0$, $\Delta_3$ is
negative ). The field $E_1$ is turned off, the initial population
of level $n$ equals 1, $g_{nn}^0=3.61$, $g_{ff}^0=9.61$,
$g_{nf}^0=5.89$, the pulse durations are equal and much shorter
than the relaxation rates, $\Omega_{nf}=0$, and the Fano
parameters are $q_{nn}=0.2$, $q_{ff}=-0.5$, $q_{nf}=10$. }
\end{figure}

\begin{figure}[!h]
\begin{center}
{\bf(a)}\hspace{80mm} {\bf(b)}\\
\includegraphics[width=0.3\textwidth]{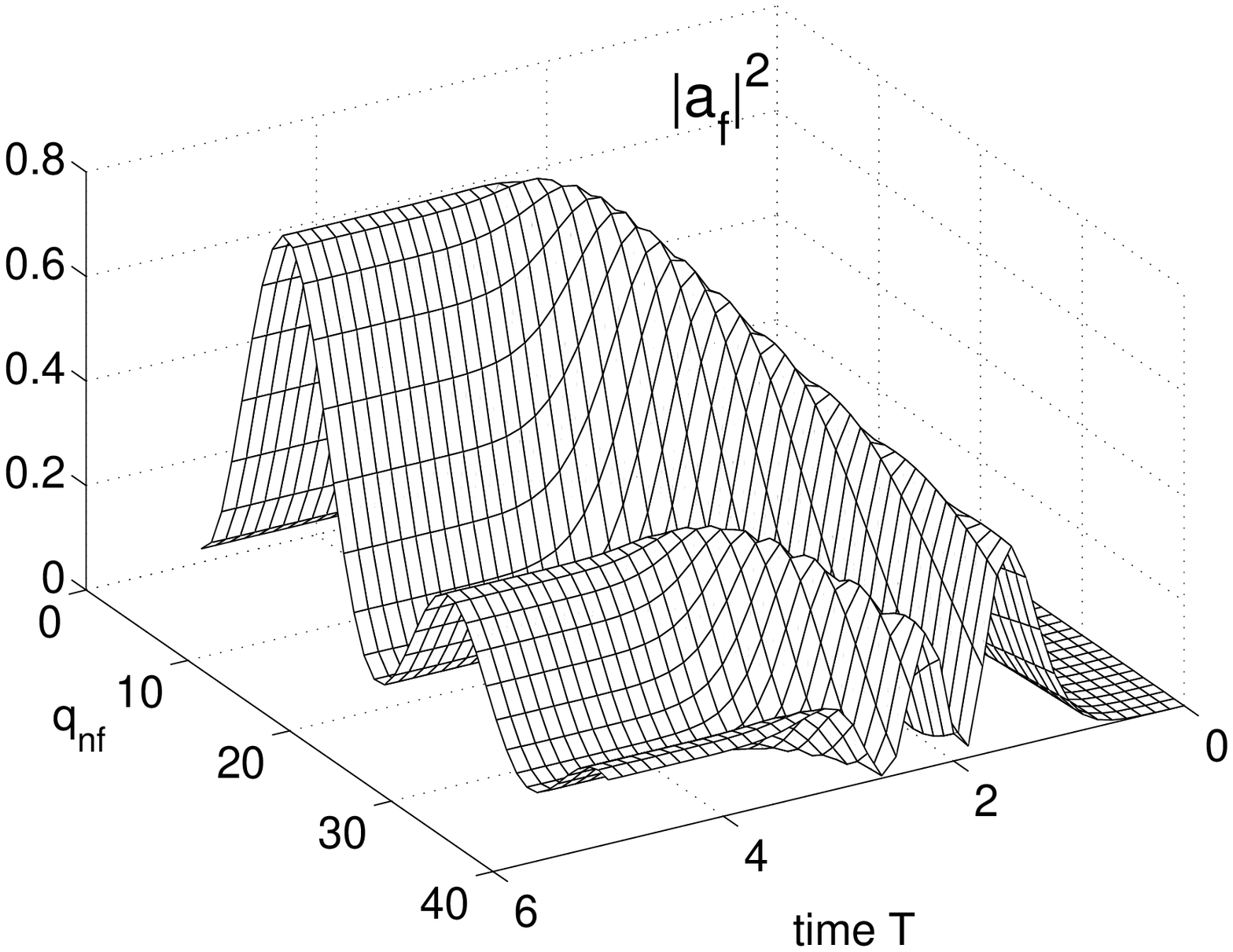}\hspace{5mm}
\includegraphics[width=0.3\textwidth]{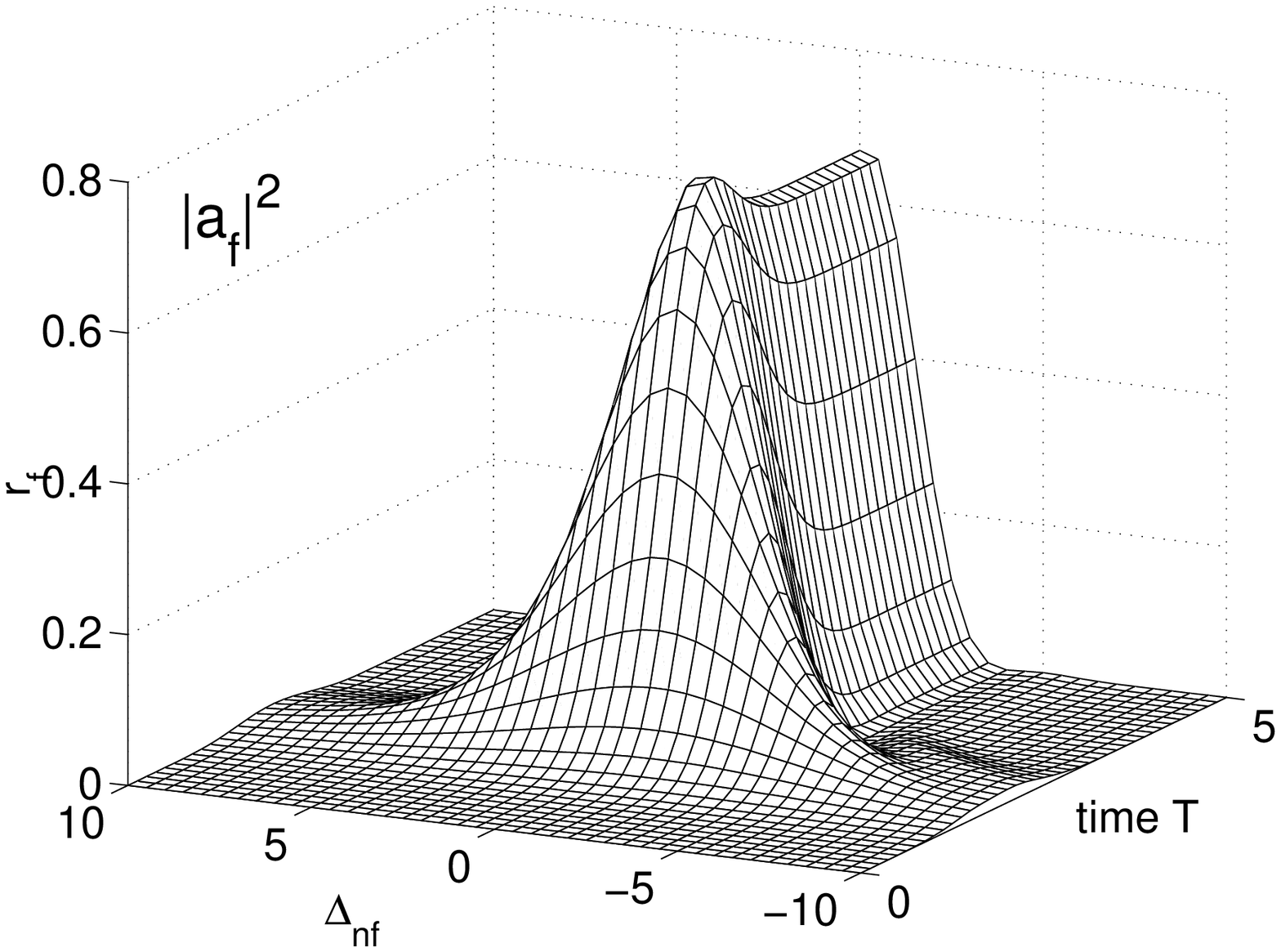}\\
{\bf(c)}\hspace{80mm} {\bf(d)}\\
\includegraphics[width=0.3\textwidth]{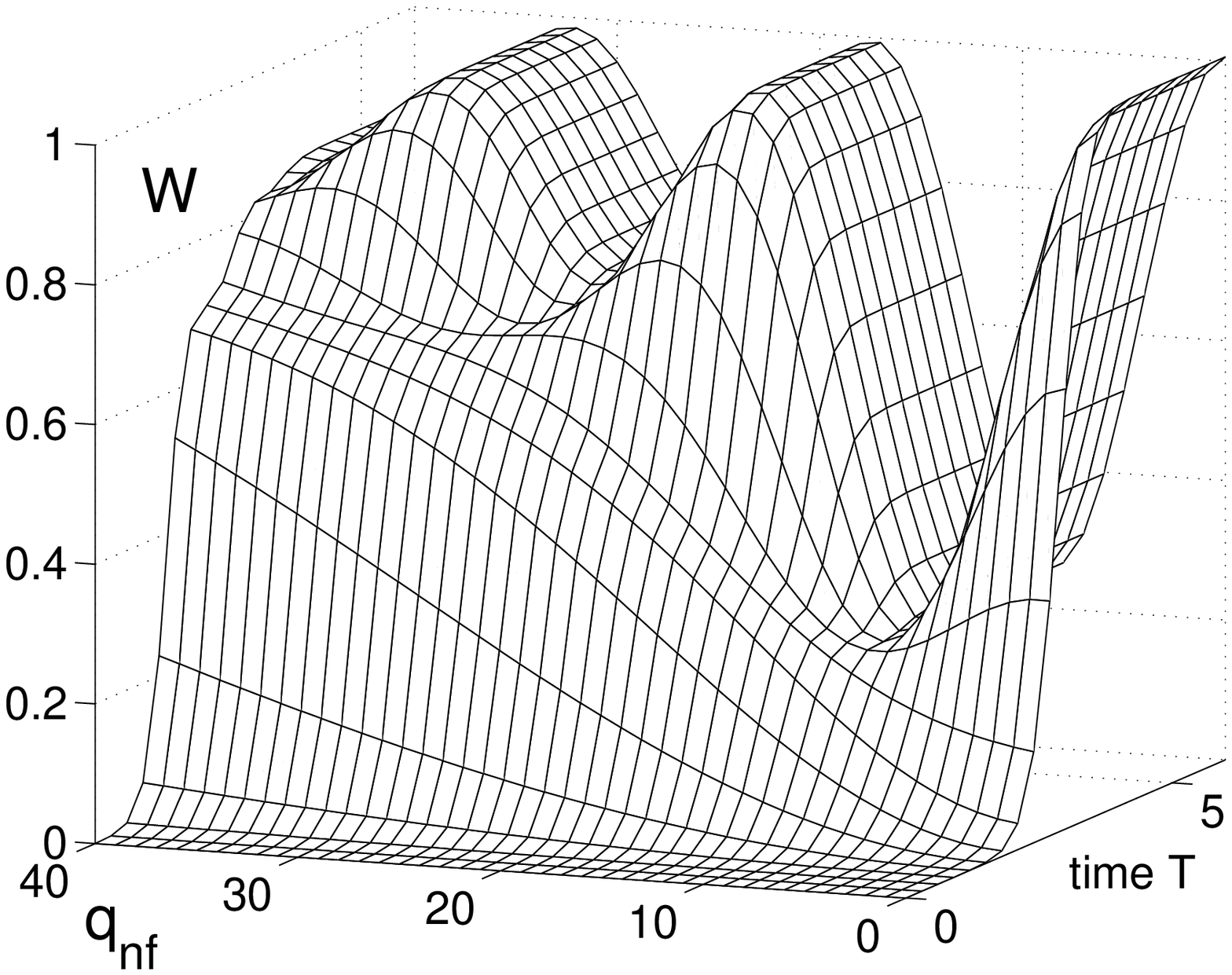}\hspace{5mm}
\includegraphics[width=0.3\textwidth]{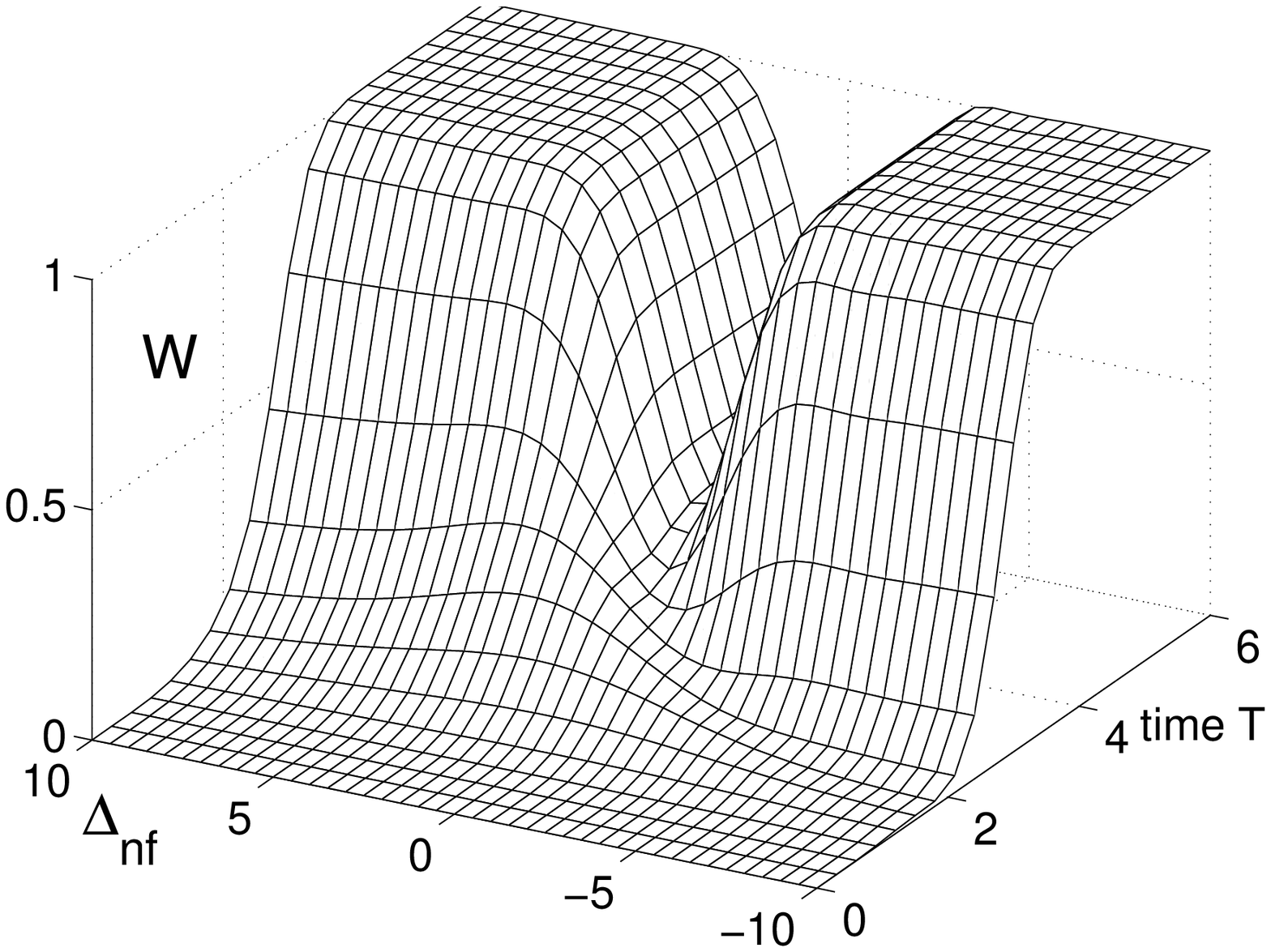}
\end{center}
 \caption{\label{p02} Dependence of the population
transfer (a,b) and dissociation (c,d) on the Fano parameter
$q_{nf}$ (a,c) and on two-photon detuning $\Delta_{nf}$ (b,d).
$\Delta_2=0$, $\Delta_3=-3.9$ for all plots. (a,c)
$\Delta_{nf}=0$, (b,d) $q_{nf}=10$. All other parameters are the
same as in Fig. \ref{p01}.}
\end{figure}

\begin{figure}[!h]
\begin{center}
{\bf(a)}\hspace{80mm} {\bf(b)}\\
\includegraphics[width=0.3\textwidth]{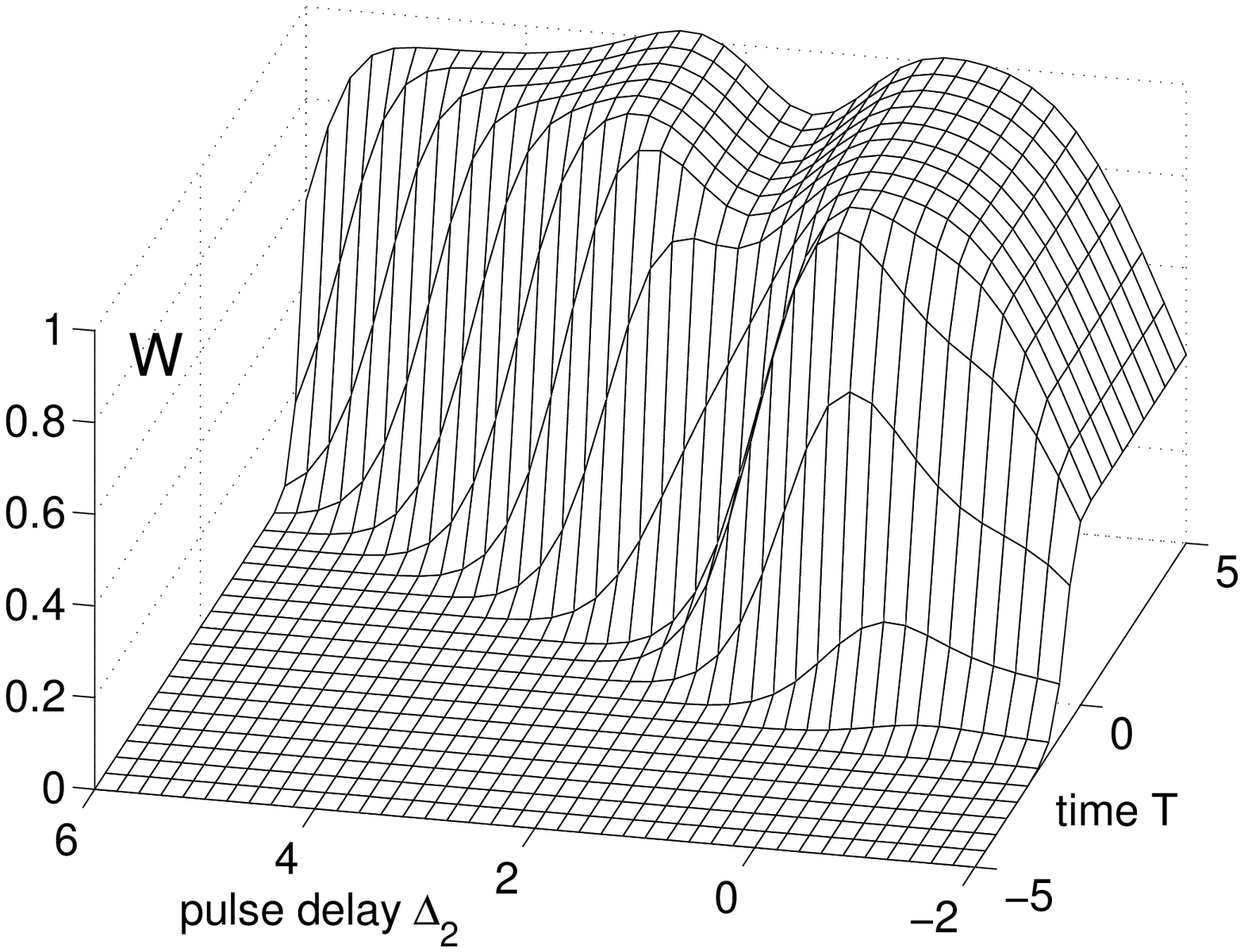}\hspace{5mm}
\includegraphics[width=0.3\textwidth]{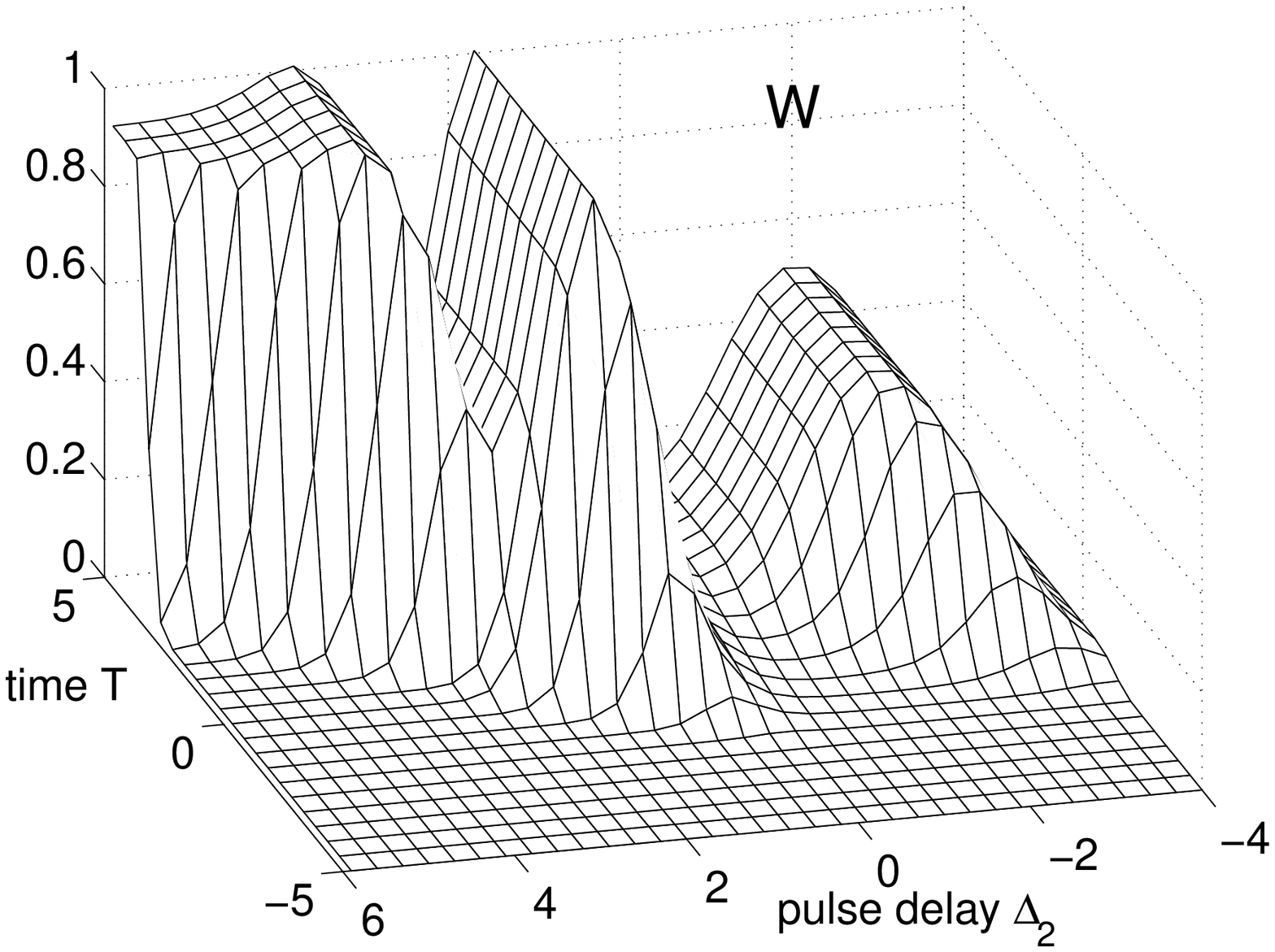}\\
{\bf(c)}\hspace{80mm} {\bf(d)}\\
\includegraphics[width=0.3\textwidth]{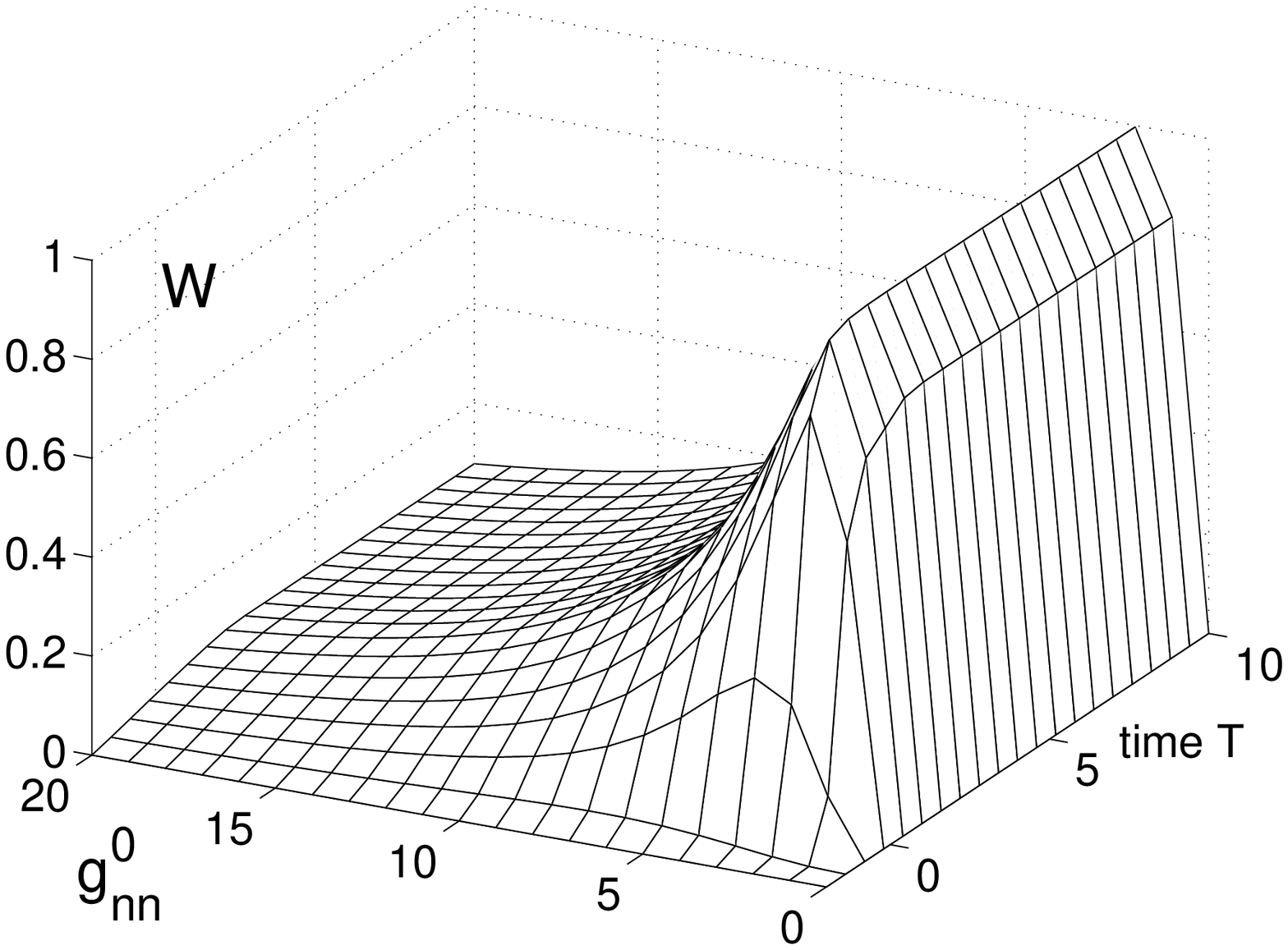}\hspace{5mm}
\includegraphics[width=0.3\textwidth]{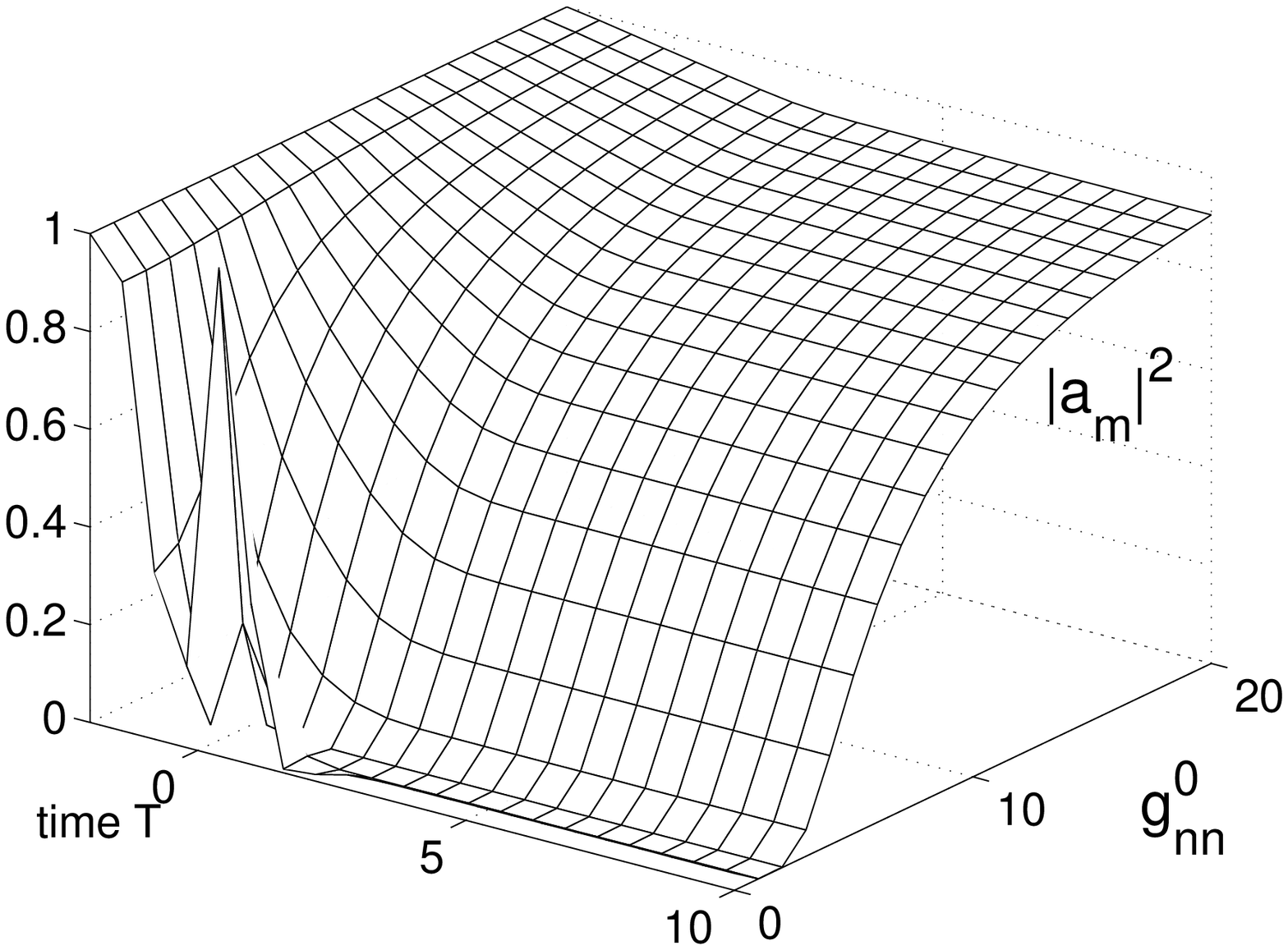}
\end{center}
\caption{\label{p03} Dependence of the two-photon dissociation
($E_3=0$) (a,b,c) and population of the ground level (d) on the
intensity and pulse delay of the driving fields. $\Omega_{mn}=0$,
$g_{mn}^0=2$. (a) $g_{nn}^0=3.61$; (b) $g_{nn}^0=400$; (c,d)
$\Delta_2=0$. All other parameters are the same as in
Fig.~\ref{p01}.}
\end{figure}

\begin{figure}[!h]
\begin{center}
{(a)}\hspace{80mm} {(b)}
\includegraphics[width=0.49\textwidth]{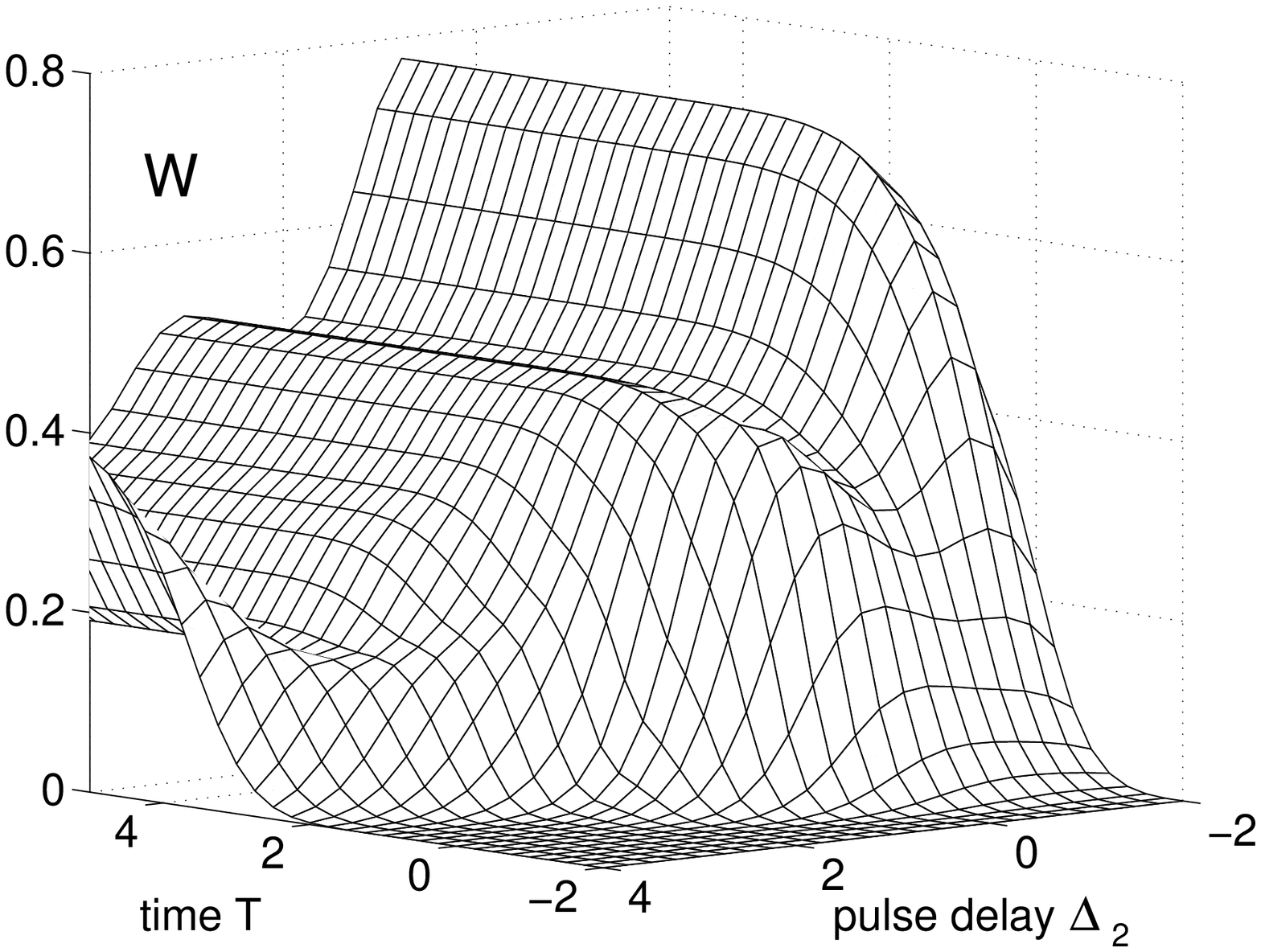}
\includegraphics[width=0.49\textwidth]{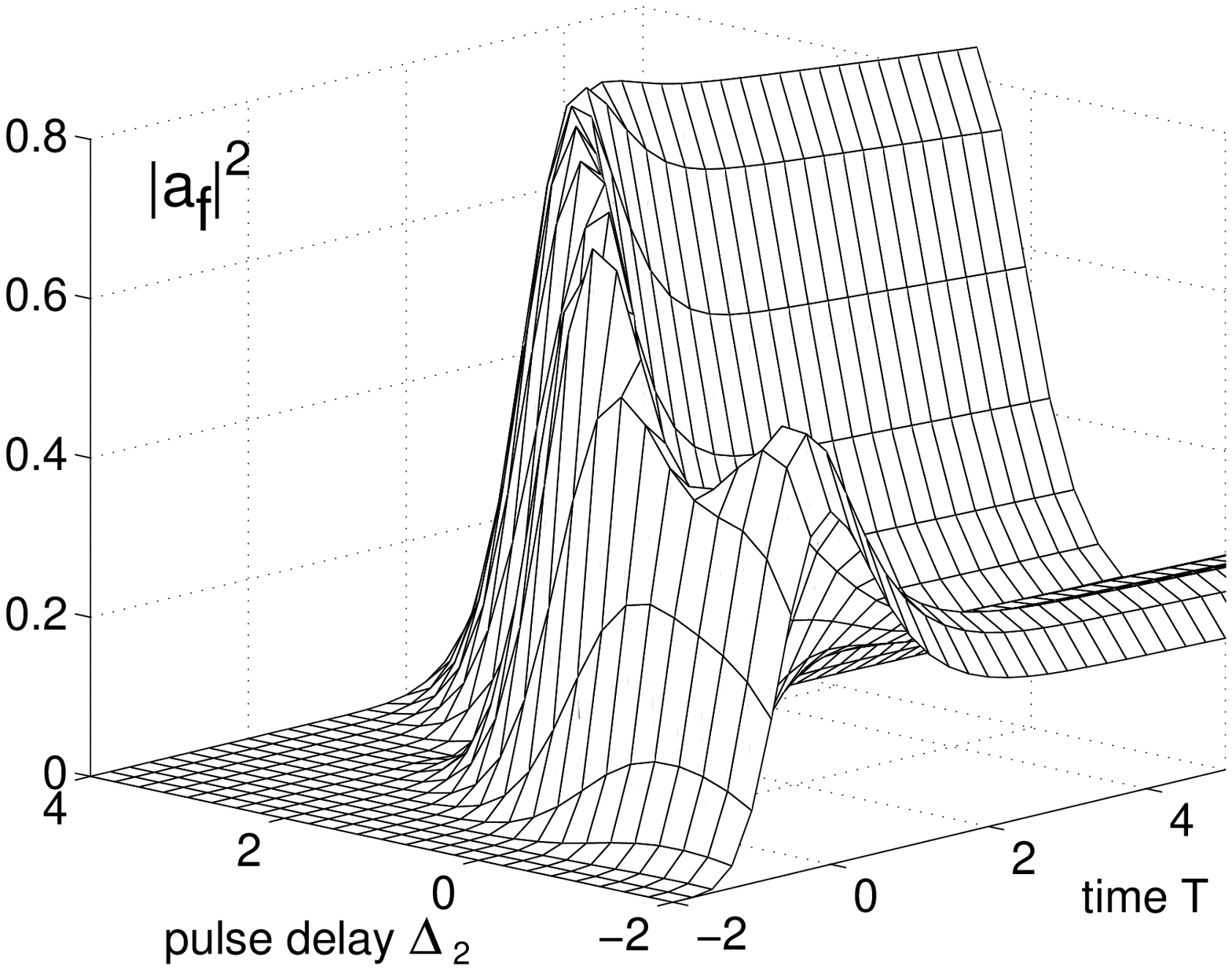}
\end{center}
\begin{center}
{(c)}\hspace{80mm} {(d)}
\includegraphics[width=0.49\textwidth]{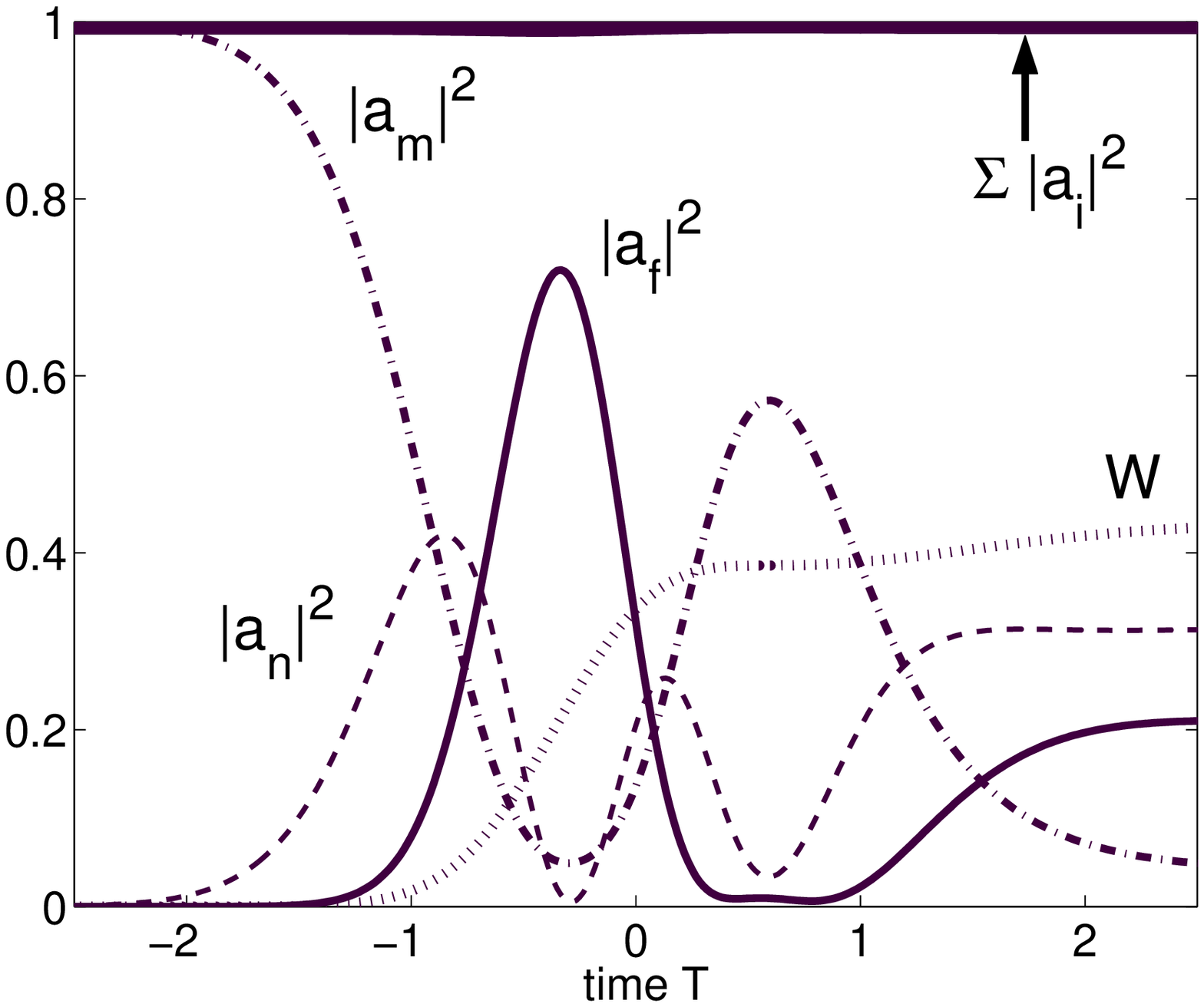}
\includegraphics[width=0.49\textwidth]{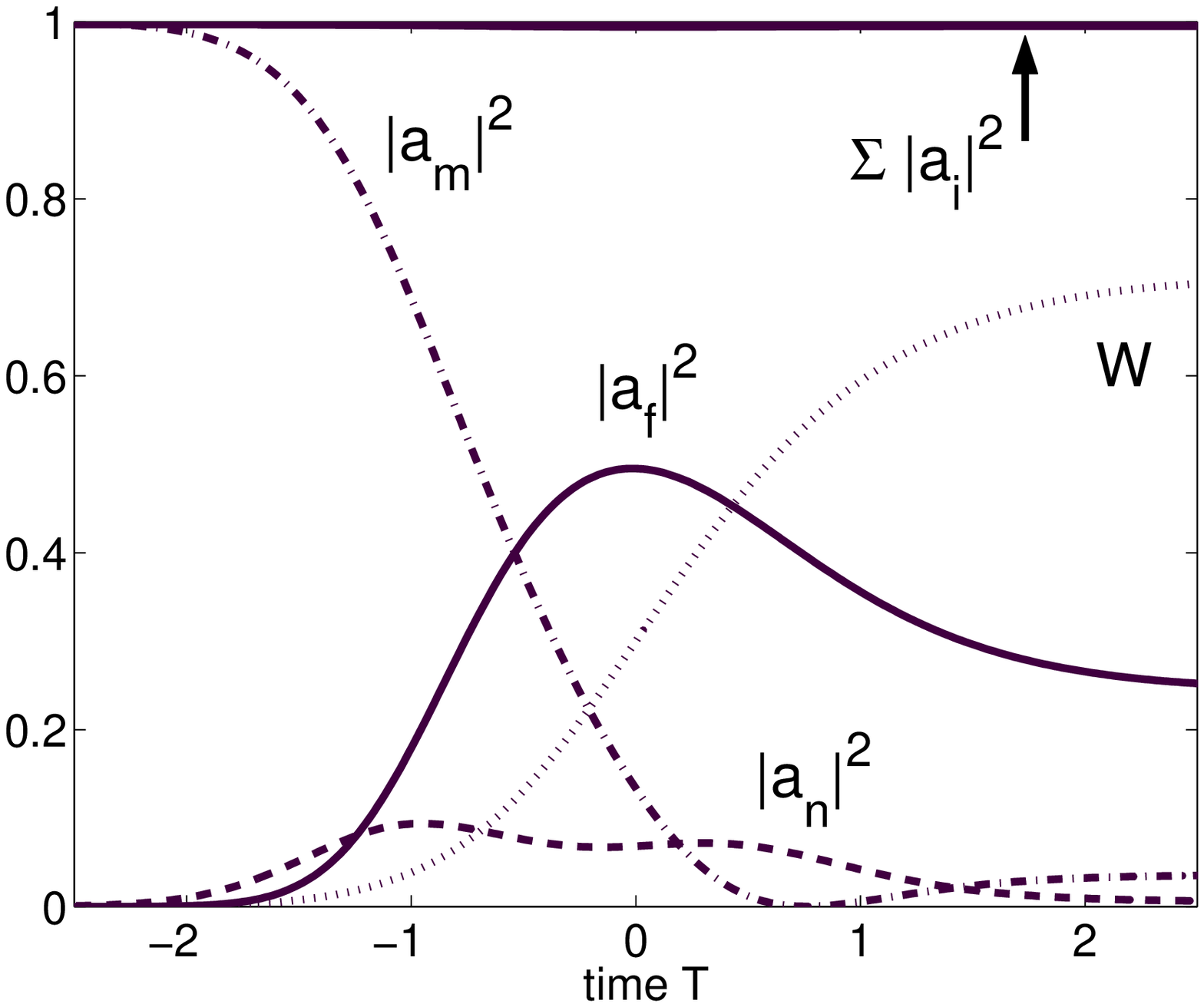}
\end{center}
\caption{\label{p1} Dissociation and population of levels
 driven by three laser pulses.  $\Omega_{mn}=\Omega_{nf}=0$,
$g_{mn}^0=2$, $g_{nn}^0=0.25$, $g_{ff}^0=0.36$, $g_{nf}^0=0.3$,
$d_2=1$, $d_3=1.6$; $\Delta_3=0$. The Fano parameters are
$q_{nn}=0.2$, $q_{ff}=-0.5$, $q_{nf}=10$. Initially, only level
$m$ is populated. (a) and (b) -- dependence on the pulse delay.
(c) and (d) -- dynamics of the populations and dissociation
($\Sigma |a_j|^2$ is the sum of populations including the energy
integrated population of the continuum $W$). (c) $\Delta_2=0$, (d)
 $\Delta_2=-1.5$.}
\end{figure}

\begin{figure}[!h]
\begin{center}
{(a)}\hspace{80mm} {(b)}
\includegraphics[width=0.49\textwidth]{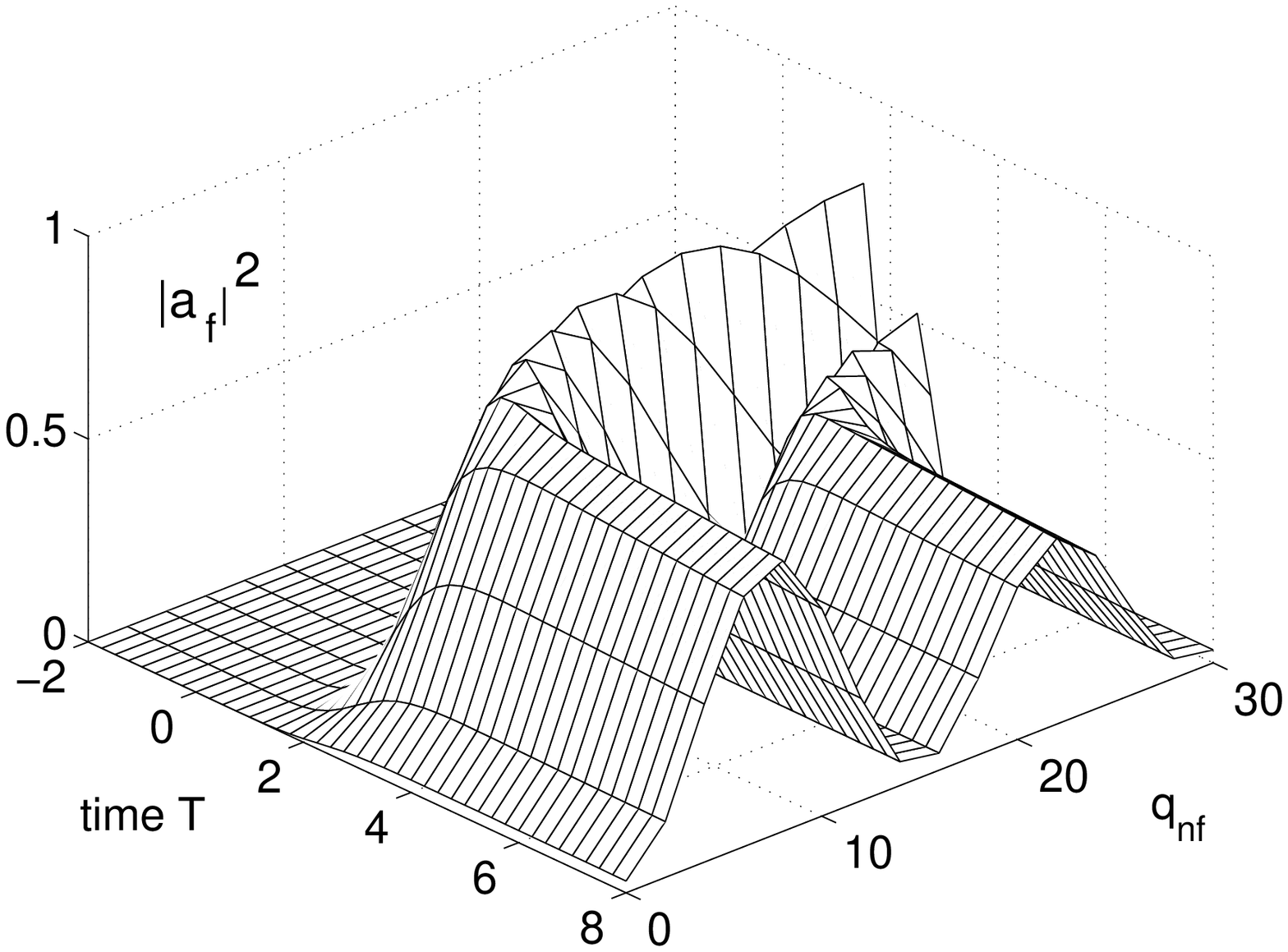}
\includegraphics[width=0.49\textwidth]{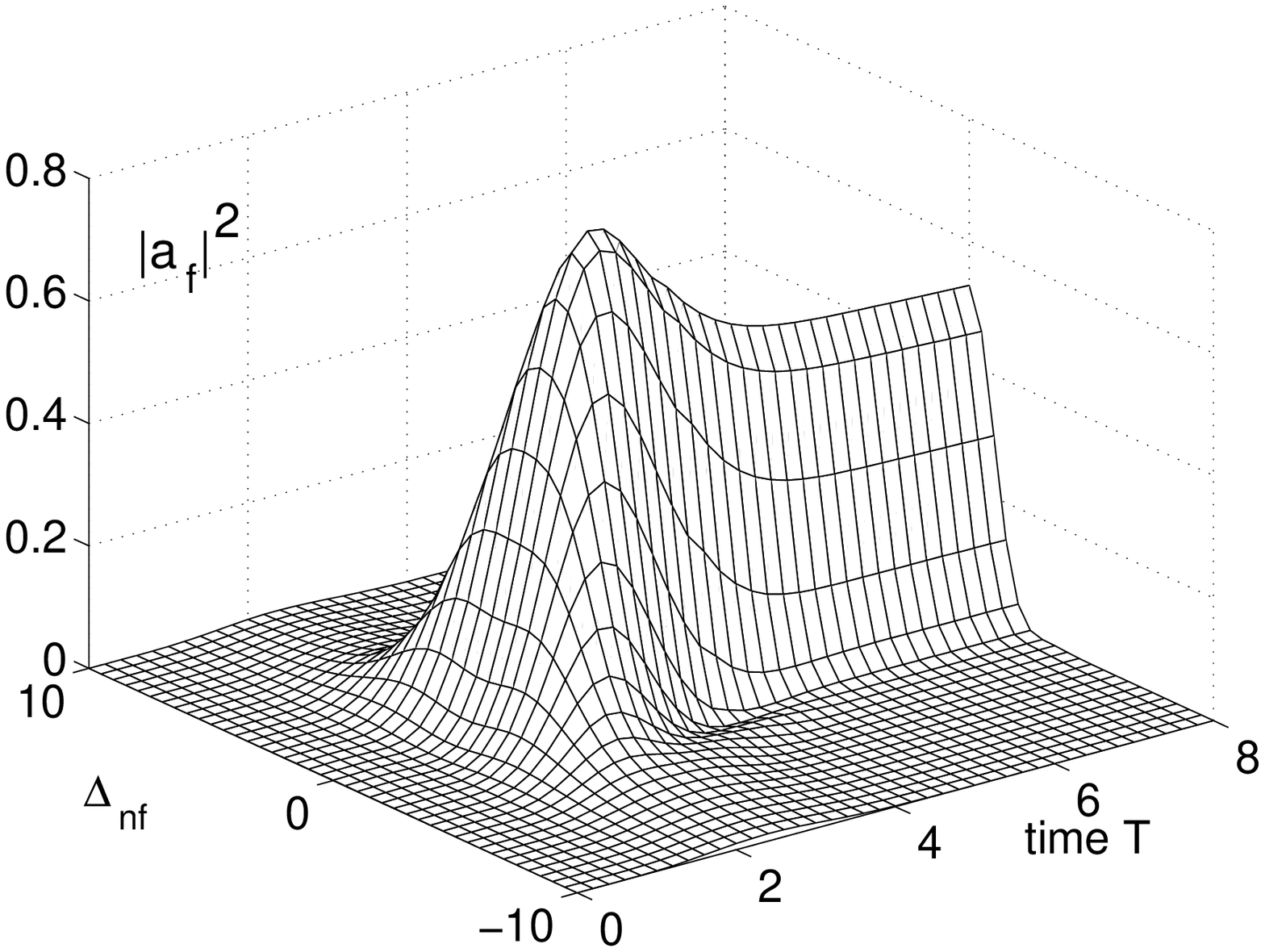}
\end{center}
\vspace{-5mm} \caption{\label{p2}  Dependence of the three-pulse
driven population of level $f$ on the Fano parameter $q_{nf}$ (a)
and on the two-photon detuning $\Delta_{nf}$ (b). $\Delta_2=2.8$,
$\Delta_3=0$. (a) $\Omega_{nf}=0$, (b)  $q_{nf}=10$. All other
parameters are the same as in the previous figure.}
\end{figure}

\begin{figure}[!h]
\begin{center}
{(a)}\hspace{80mm} {(b)}
\includegraphics[width=0.49\textwidth]{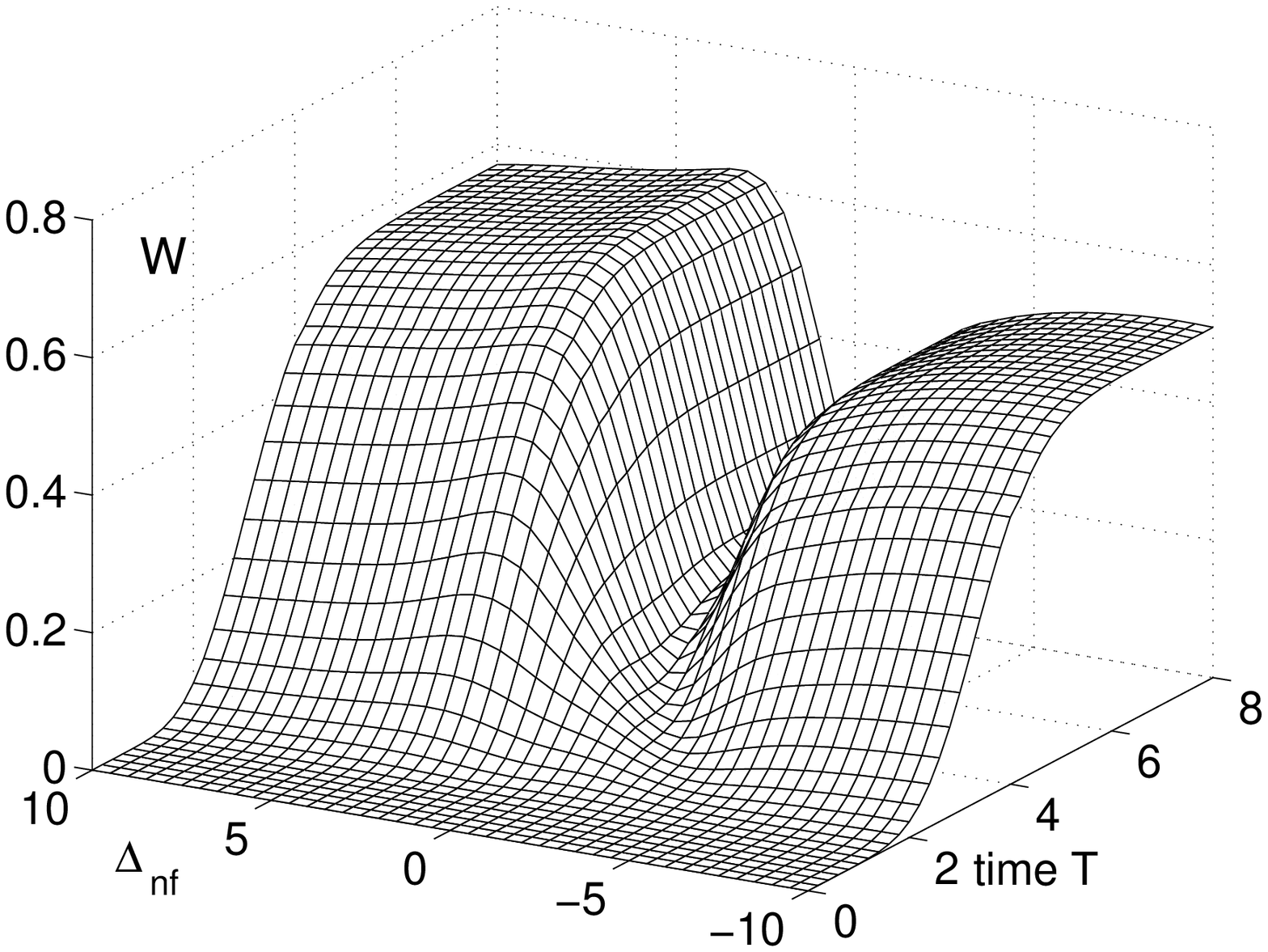}
\includegraphics[width=0.49\textwidth]{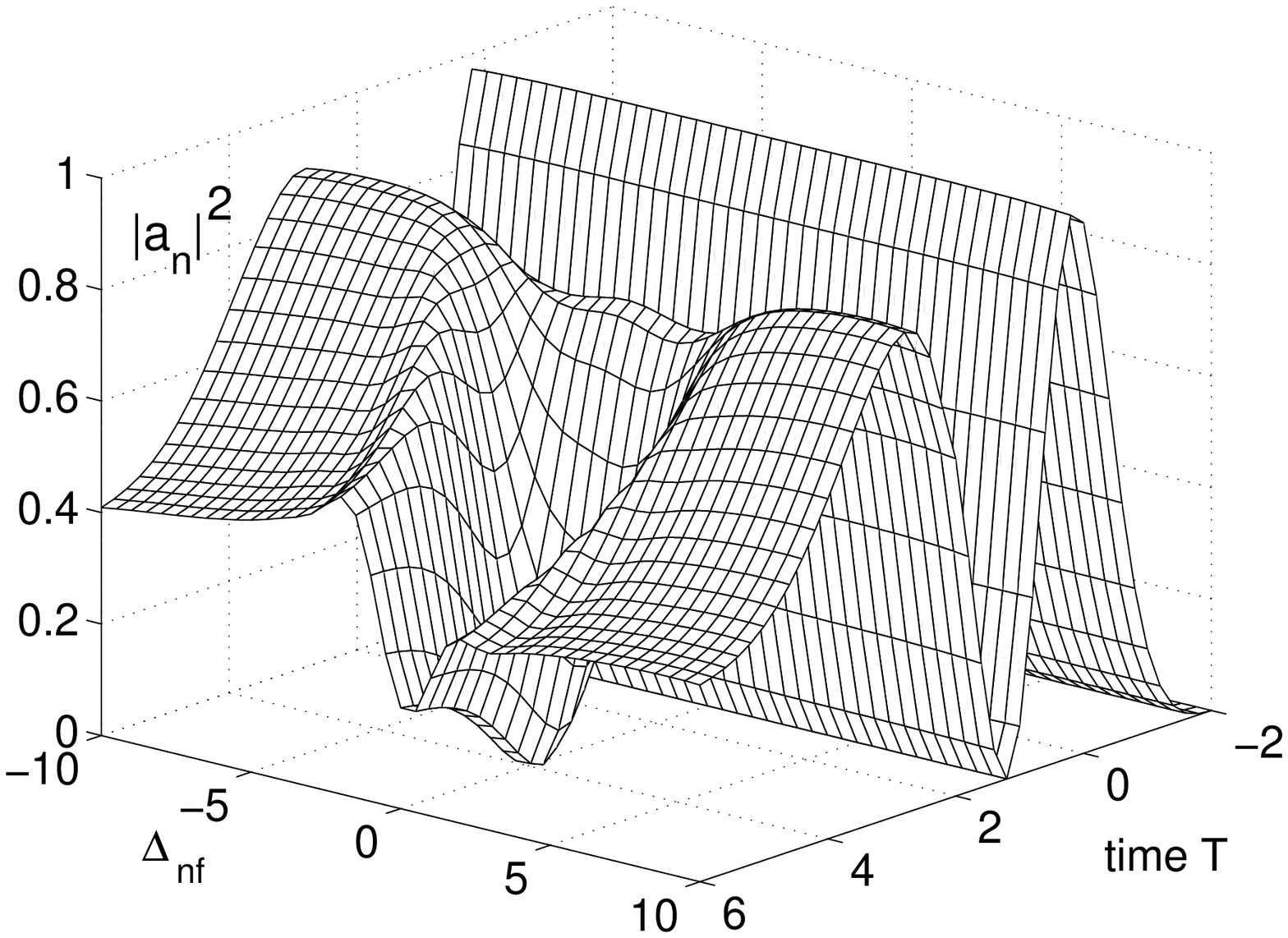}
\end{center}
\vspace{-5mm} \caption{\label{p3} Dependence of the dissociation
and of the population of the intermediate level $n$ on the
two-photon detuning, where $q_{nf}=10$.  All other parameters are
the same as in the previous figure.}
\end{figure}


\subsection{Suppression of photodissociation by continuum coherence
and overlap of two LICS}\label{2l} Figures \ref{p01} and \ref{p02}
demonstrate feasibilities of manipulating constructive and
destructive interference aimed at control of the population
transfer between the upper levels $n$ and $f$. Here the first
field is turned off and only level $n$ is initially populated.
Figure \ref{p01} simulates the population dynamics when the
carrier frequencies of the fields are tuned to the two-photon
resonance with $\omega_{fn}$. It shows a dramatic dependence of
the output on the temporal shift between the pulse peaks. Such
dynamics is stipulated by the time-dependent shift of the
two-photon resonance, which determines constructive or,
alternatively, destructive interference. In other words, the
position, strength and overlap the two LICS  vary in time, which
bring about the dynamics and consequences of the interference. For
the given parameters,  we have almost complete suppression of
dissociation [plot (a)] and about 80\% population transfer to
level $f$ [plot (b)]. This occurs for counterintuitive order of
pulses, when third field reaches its maximum at about four pulse
half-duration \textit {before} the maximum of second pulse.
Alternatively, almost complete photodissociation can be ensured
with different pulse shifts. Thus our simulations demonstrate a
behavior, which is in a good agreement with the conclusions and
observations of other studies \cite{Half,Yats}. Figure \ref{p02}
[plots (a) and (c)] displays a strong dependence of dissociation
and population transfer on the effective Fano parameter $q_{nf}$.
This because they determine the magnitude and sign of the
resonance shift (position and shape of LICS).   The plots (b) and
(d) show that similar manipulations can be performed by an
appropriate adjustment of the two-photon detuning.

\subsection{Photodissociation controlled by two ordered
pulses}\label{pd} Figure \ref{p03} shows the features of
dissociation from the lowest state controlled with two pulses.
Here the third field is turned off, only level $m$ is initially
populated, the carrier frequency of the first field is tuned to
resonance with the transition $mn$, and dissociation is caused by
the interference of two-step and two-photon processes. The plots
indicate that, due to the time-dependent shift of one-photon
resonance induced by the second field, the dynamics of
photodissociation and its dependence on the pulse sequences and
shift changes considerably with the increase of the peak intensity
of the second field [plots (a) and (b)]. At a given pulse-shift
there exists an optimal intensity of this field [plot (c)].  The
increase of the intensity of the second field leads to decoupling
of the first one with the transition $mn$ [plot (d)] and,
consequently, to a decrease in the overall dissociation [plot
(c)]. Rabi oscillations of the populations at the transition $mn$
in the range of relatively small intensities of the second field
[plot (d)] may give rise to substantial modification of the
dynamics and to suppression of dissociation [plot (b)].

\subsection{Three-pulse control of photodissociation and
population transfer by continuum coherence}\label{tp} With an
understanding of the features described above, we shall
investigate the dynamics of more complex scheme  which embraces
both schemes considered above by  assuming all three fields at
$\omega_1$, $\omega_2$ and $\omega_3$ (Fig.~\ref{lev}) turned on.
First, we shall simulate the case when one-, two- and three-photon
resonances are fulfilled ($\Omega_{mn}=\Omega_{nf}=0$), and the
first and second pulses are peaked at the same instant as the
third pulse, which is somewhat longer. The second pulse is of the
same duration as the first one and can be either advanced or
delayed with respect to the other two.

Figure~\ref{p1} [plots (a) and (b)] shows that the dissociation
output and population of the upper level can be controlled within
the wide intervals by adjustment of the sequence and the time
shift of the second pulse. With two examples [plots (c) and (d)],
the simulation demonstrates the feasibility of achieving either
comparable populations of all three discrete and integrated
continuum  states by the end of the pulses [plots (c), where all
pulses peak at the same instant] or, alternatively, of ensuring a
dominant population of the upper discrete state and dissociation
output by advancing the second pulse regarding the  other two
[plot (d)]. Here, $\Sigma |a_j|^2$ is the sum of populations
including the energy integrated population of the continuum $W$.
Figure \ref{p2} [plot (a)] shows that the dynamics of the
populations is considerably different for different Fano
parameters (keeping the other parameters the same). It also shows
that, for fixed time delays, the output can be controlled by the
adjustment of appropriate two-photon detuning [plot (b)]. Figure
\ref{p3} shows similar feasibilities of manipulation of the
populations of the intermediate level and dissociation. Dark
resonances, characterized by  suppressed dissociation and
selective population redistribution between the coupled bound
states, or, alternatively, enhanced dissociation  are shown to be
achieved by variation of two-photon (and consequently
three-photon) detuning.

\section{Conclusion}
The theory of adiabatic passage between three discrete levels and
a low-lying dissociation continuum is developed. Possible quantum
control of two-photon dissociation ($\Lambda$-scheme) using
auxiliary short laser pulses at adjacent bound-free transitions
($V$-configuration) is shown. The proposed three-pulse scheme
enables one to manipulate dissociation through the dark states not
connected with the ground one by the allowed transition. Besides
dissociation, the scheme under investigation enables  one to
control the population transfer between two upper discrete levels
via the lower-energy dissociation continuum, while a direct
transition between these states is not allowed.  A system of
coupled equations for the probability amplitudes is employed,
where three time-shifted pulses drive strongly the quantum system,
so that nonlinear quantum interference occurs through the overlap
of two laser-induced continuum structures. The opportunities of
manipulating photoinduced processes by judiciously
frequency-detuned and time-delayed pulses are explored through
extensive numerical simulations with the aid of a model relevant
to typical experiments. The simulations display good agreement
with the known experiments on two-pulse population transfer
between metastable states via the ionization continuum and with
experiments on two-photon photodissociation. The dependence of the
dissociation spectrum on the magnitudes of the composite Fano
parameters for excited levels, on detuning between two
laser-induced continuum structures, on intensities, and on the
counterintuitive order and delay of the pulses  are investigated.
We have demonstrated the opportunities of considerable suppression
of photodissociation  and considerable transfer of the lower-state
population to the excited states or, alternatively, almost
complete photodissociation, or redistribution the population
between continuous and discrete states with the required ratio by
the end of the pulses. The gained efficiency and flexibility may
compensate for the increase in experimental complexity.
\begin{acknowledgments}
This work has been supported in part by the International
Association (INTAS) of the European Community for the promotion of
cooperation with scientists from the New Independent States of the
former Soviet Union (grant INTAS-99-00019). The authors thank
K.~Bergmann for useful, encouraging discussions.
\end{acknowledgments}

\begin{thebibliography}{99}

\bibitem{TR1} D. J. Tennor and  S. A. Rice, "Control of Selectivity of Chemical Reaction via
Control of Wave Packet Evolution," J. Chem. Phys. \textbf{ 83},
5013-18 (1985).

\bibitem{BrSh}M. Shapiro and P. Brumer, "Laser Control of Product Quantum State Populations in
Unimolecular Reactions," J. Chem. Phys. \textbf{ 84}, 4103-04
(1986).

\bibitem{TR2} D. J. Tennor, R. Kosloff, and S. A. Rice, "Coherent Pulse Sequence
Induced Control of Selectivity of Reactions - Exact
Quantum-Mechanical Calculations," J. Chem. Phys. \textbf{ 85},
5805-20 (1986).

\bibitem{Br1} T. Seideman, M. Shapiro, and P. Brumer, "Coherent Radiative Control of
Unimolecular Reactions: Selective Bond Breaking with Picosecond
Pulses," J. Chem. Phys. \textbf{ 90}, 7132-36 (1989).

\bibitem{Sh1} P. Brumer and M. Shapiro, "Laser Control of Molecular
Processes," Ann. Rev. Phys. Chem. \textbf{ 43}, 257-82 (1992).


\bibitem{Br2} A. Shnitman, I. Sofer, I. Golub, A. Yogev, M. Shapiro, Z. Chen, and P.
Brumer, "Experimental Observation of Laser Control: Electronic
Branching in the Photodissociation of Na$_2$," Phys. Rev. Lett.
\textbf{ 76}, 2886-89 (1996).

\bibitem{R1} R. J. Gordon and S. A. Rice, "Active Control of the Dynamics of Atoms and
Molecules," Ann. Rev. Phys. Chem. \textbf{ 48}, 601-41 (1997).

\bibitem{Br3} M. Shapiro and P. Brumer, "Quantum Control of Chemical Reactions," Trans.
Faraday Soc. \textbf{ 93}, 1263-77 (1997).

\bibitem{Br4} M. Shapiro, P. Brumer, and Z. Chen, "Simultaneous Control of Selectivity and
Yield of Molecular Dissociation: Pulsed Incoherent Interference
Control," Chem. Phys. \textbf{ 217}, 325-340 (1997).

\bibitem{Br5} P. Brumer and M. Shapiro, "Quantum Interference in the Control of Molecular
Processes", Phil. Trans. R. Soc. London A \textbf{ 355}, 2409-12
(1997).

\bibitem{R4} M. N. Kobrak and S. A. Rice, "Selective Photochemistry via Adiabatic
Passage: An Extension of Stimulated Raman Adiabatic Passage for
Degenerate Final States," Phys. Rev. A \textbf{ 57}, 2885-94
(1998).

\bibitem{Gor1} R. J. Gordon, L. Zhu, and T. Seideman, "Coherent Control of Chemical
Reactions," Acc. Chem. Res.  \textbf{ 32}, 1007-16 (1999).

\bibitem{Sh2} E. Frishman, M. Shapiro, and P. Brumer, "Coherent Enhancement and Suppression
of Reactive Scattering and Tunneling," J. Chem. Phys. \textbf{
110}, 9-11 (1999).

\bibitem{Sh3} M. Shapiro and P. Brumer, "Coherent Control of Atomic, Molecular and
Electronic Processes," in \textit{ Advances in Atomic, Molecular
and Optical Physics}, Vol. \textbf{ 42}, edited by B. Bederson and
H. Walther (Academic Press, San Diego, 1999), pp. 287-343.

\bibitem{Mcool} D. J. Tennor, R. Kosloff, and A. Bartanab, "Laser Cooling of
Internal Degrees of Freedom of Molecules by Dynamically Trapped
States," Faraday Discuss. \textbf{ 113}, 365-83 (1999).

\bibitem{R5} S. A. Rice, "Active Control of Molecular Dynamics: Coherence versus Chaos,"
J. Stat. Phys. \textbf{ 101}, 187-212 (2000).

\bibitem{R6} S. A. Rice, "Molecular Dynamics: Optical Control of Reactions," Nature \textbf{
403}, 496-7 (2000).

\bibitem{R9} S. A. Rice and M. Zhao, \textit{ Optical Control of Molecular Dynamics}
(Wiley, New York, 2000).

\bibitem{R8} V. Kurkal and S. A. Rice, "Sequential STIRAP-Based Control of the HCN-CNH
Isomerization," Chem. Phys. Lett. \textbf{ 344}, 125-37 (2001).



\bibitem{Fan61} U. Fano, "Effects of Configuration Interaction on Intensities
and Phase Shifts,"  Phys. Rev. \textbf{ 124}, 1866-78 (1961).

\bibitem{FanCo} U. Fano and J. V. Cooper, "Spectral Distribution of Atomic
Oscillator Strengths," Rev. Mod. Phys., \textbf{ 40}, 441-507
(1968).


\bibitem{Fene61} L. Armstrong, Jr., B. L. Beers and S. Feneuille,
"Resonant Multiphoton Ionization via the Fano Autoionization
Formalism," Phys. Rev. A \textbf{ 12}, 1903-10 (1975).

\bibitem{GP1} Yu. I. Heller and A. K. Popov, "Autoionizing-Like Resonances Induced by
Laser Field," {Opt. Commun.} \textbf{ 18}, 7-8 (1976).

\bibitem{GP2} Yu. I. Heller and A. K. Popov, "Parametric Generation and Absorption of
VUV Radiation Controlled by Laser-Induced Autoionizing-Like
Resonances in the Continuum", Opt. Commun. \textbf{ 18}, 449-51
(1976).

\bibitem{GP3} Yu. I. Heller and A. K. Popov, "Laser-Induced Narrowing of Autoionizing
Resonances Studied by the Method of Parametric Generation," Phys.
Lett. \textbf{ 56A}, 453-4 (1976).

\bibitem{GP4} Yu. I. Heller and A. K. Popov, "On Inducing Narrow Nonlinear Resonances
in the Continuum," Kvantovaya Elektronika \textbf{ 3}, 1129-31
(1976) [English translation:  Soviet Journal of Quantum
Electronics].

\bibitem{GP5} A. K. Popov and Yu. I. Heller, "Autoionizing Levels in Four-wave
Spectroscopy," in \textit{ Applied Spectroscopy} (AN SSSR, Moscow,
1977), pp. 50-53 (in Russian).

\bibitem{GP6} Yu. I. Heller and A. K. Popov, "Nonlinear Polarization Resonances in the
Continuum," J. Exper. Theor. Phys. \textbf{ 51},  255-9 (1980)
[translated from Zhurnal Eksperimental'noi i Teoreticheskoi Fiziki
\textbf{ 78},  506-15 (1980)].

\bibitem{Gel} Yu.I. Heller and A.K. Popov, \textit{ Laser Induction
of Nonlinear Resonanses in Continuous Spectra} (Novosibirsk,
Nauka, 1981) [Engl. translation: {J. Sov. Laser Research} \textbf{
6}, N 1-2 (1985), Plenum, c/b Consultants Bureau, NY, USA].

\bibitem{GP7} Yu. I. Heller, V. V. Lukinykh, A. K. Popov, and V. V. Slabko,
"Experimental Evidence for Laser-Induced Autoionizing-Like
Resonances in the Continuum," {Pisma v Zhurnal Tekhnicheskoi
Fiziki} \textbf{ 6},  151-5 (1980) [English translation: J. Techn.
Phys.  Lett.].

\bibitem{GP8} A. K. Popov, Yu. I Heller, V. F. Lukinykh, and V. V. Slabko,
"Autoionizing-Like Resonances Induced by a Laser Field in the
Spectral Continuum of CsI," Proceedings of the International
Conference LASERS'80, New Orleans, Louisiana, (STS Press, McLean,
VA, 1981), pp. 735-40.

\bibitem{GP9} Yu. I. Heller, V. F. Lukinykh, A. K. Popov, and V. V. Slabko,
"Experimental Evidence for Laser-Induced Autoionizing-Like
Resonances in the Continuum," Phys. Lett.  \textbf{ 82A} 4-6
(1981).

\bibitem{GP10} Yu. I. Heller, V. V. Lukinykh, A. K. Popov, and V. V. Slabko,
"Autoionizing-Like Resonances, Induced in the Continuum Spectrum
of  Atomic Cesium," Optika i Spektroskopiya \textbf{ 51}, 732-4
(1981) [English translation: Optics and Spectroscopy (USSR)].

\bibitem{GP11} Yu. I. Heller and A. K. Popov, "Laser-Induced Narrowing of Autoionizing
Resonances in Multiphoton Ionization Spectrum,"  Opt. Commun.
\textbf{ 38}, 345-7 (1981).

\bibitem{Zol} P. Lambropoulos and P. Zoller, "Autoionizing States in Strong
Laser Fields,"  Phys. Rev. A \textbf{ 24}, 379-97 (1981).

\bibitem{Rza81} K. Rzazewski and J. H. Eberly, "Confluence of Bound-Free
Coherences in Laser-Induced Autoionization,"  Phys. Rev. Lett.
\textbf{ 47}, 408-12 (1981).

\bibitem{Pav82} L. I. Pavlov, S. S. Dimov, D. I. Mechikov, G. M. Mileva, K. V.
Stamenov, and G. B. Altschuler, "Efficient Tunable Tripler of
Optical Frequency at an Autoionizing-Like Resonance in a
Continuum," Phys. Lett. \textbf{ 89 A}, 441-3 (1982).

\bibitem{GP12} S. S. Dimov, Yu. I. Heller, L. I. Pavlov, A. K. Popov, and K. V.
Stamenov, "Laser-Induced Nonlinear Resonances in the Continuum at
Third-Harmonic Generation in Na Vapor," Applied Physics B \textbf{
30}, 35-40 (1983).

\bibitem{Rza83} K. Rzazewski and J. H. Eberly, "Photoexcitation of an
Autoionizing Resonance in the Presence of Off-Diagonal
Relaxation,"  Phys. Rev. A \textbf{27}, 2026-42 (1983).

\bibitem{GP13} S. S. Dimov, Yu. I. Heller, L. I. Pavlov, and A. K. Popov,
"Induced Autoionizing-like Resonances in Third- and Fifth-Order
Nonlinear Susceptibilities," Sov. J. Quantum Electron. \textbf{
13}, 1075-81 (1983) [translated from Kvantovaya Elektronika
\textbf{ 10}, 1635-45 (1983)].

\bibitem{Kn84} P. L. Knight, M. A. Lauder, P. M. Radmore, and B. J. Dalton,
"Making Atoms Transparent: Trapped Superpositions," Acta Phys.
Austriaca \textbf{ 56}, 103-17 (1984).

\bibitem{KnLICS} P. L. Knight, "Laser-Induced Continuum Structure," Comments
At. Mol. Phys. \textbf{ 15}, 193-214 (1984).



\bibitem{Lam87} B.-N. Dai and P. Lambropoulos, "Laser-Induced Autoionizing-Like
Behavior, Population Trapping, and Stimulated Raman Processes in
Real Atoms,"  Phys. Rev. A \textbf{ 36}, 5205-8 (1987).

\bibitem{Hut} M. H. R. Hutchinson and K. M. M. Ness, "Laser-Induced
Continuum Structure in Xenon," Phys. Rev. Lett. \textbf{ 60},
105-7 (1988); M. H. R. Hutchinson and K. M. M. Ness, "Hutchinson
and Ness Reply," Phys. Rev. Lett. \textbf{ 62}, 112 (1989).

\bibitem{LamCom} X. Tang, A. L'Huillier, and P. Lambropoulos, "Comment on
'Laser-Induced Continuum Structure in Xenon'," Phys. Rev. Lett.
\textbf{ 62}, 111 (1989).

\bibitem{Kn} P. L. Knight, M. A. Lander, and B. J. Dalton, "Laser-Induced Continuum
Structure,"  Phys. Rep. \textbf{ 190}, 1-61 (1990).
\bibitem{Char91} Y. L. Shao, D. Charalambidis, and C. Fotakis, "Observation of
laser-induced Continuum Structure in Ionization of Sodium," Phys.
Rev. Lett. \textbf{ 67}, 3669-72 (1991).

\bibitem{Cav91} S. Cavalieri, F. S. Pavone, and M. Matera, "Observation of a
Laser-Induced Resonance in the Photoionization Spectrum of
Sodium," Phys. Rev. Lett. \textbf{ 67}, 3673-6 (1991).

\bibitem{Fau93} O. Faucher, D. Charalambidis, C. Fotakis, Jian
Zhang,
and P. Lambropoulos, "Control of Laser-Induced Continuum Structure
in the Vicinity of Autoionizing States,"  Phys. Rev. Lett.
\textbf{ 70}, 3004-7 (1993).

\bibitem{Char93} O. Faucher, Y. L. Shao, and D. Charalambidis, "Modification of
a Structured Continuum Through Coherent Interactions Observed in
Third Harmonic Generation ,"  J. Phys. B \textbf{ 26}, L309-13
(1993)



\bibitem{Fau94} O. Faucher, Y. L. Shao, D. Charalambidis, and C. Fotakis,
"Laser-induced Modification of a Structured Continuum Observed in
Ionization and Harmonic Generation," Phys. Rev. A \textbf{ 50},
641-48 (1994).

\bibitem{CavJ95} S. Cavaliery, R. Eramo, and L. Fini, "Laser-induced
Structure in the Continuum of Sodium: A Weak Dressing Field
Measurement," J. Phys. B: At. Mol. Opt. Phys. \textbf{ 28},
1739-1801 (1995).

\bibitem{Cav95} S. Cavalieri, R. Eramo, R. Buffa, and M. Matera,
"Laser-induced Autoionizing and Continuum Structures: Line-Shape
Study in the Presence of Continuum-Continuum Transitions," Phys.
Rev. A \textbf{ 51}, 2974-81 (1995).


\bibitem{bull} A. K. Popov, "Inversionless
Amplification and Laser-Induced Transparency at Discrete
Transitions and Transitions to the Continuum" (review), {Bull.
Russ. Acad. Sci. (Physics)} \textbf{ 60}, 927-45 (1996);
\url{http://xxx.lanl.gov/abs/quant-ph/0005108}.

\bibitem{Car} C. E. Carroll and F. T. Hioe, "Selective Excitation and Structure
in the Continuum," Phys. Rev. A \textbf{ 54}, 5147-51 (1996).

\bibitem{A2}  A. D. Gazazyan and R. G. Unanyan, "Laser Velocity Selection of Atomic
Beams through the Formation of Laser-Induced Continuum with a
Structure," Laser Physics \textbf{ 6}, 946-48 (1996).


\bibitem{Cav97} R. Eramo, S. Cavalieri, L. Fini, M. Matera, and L. F. DiMauro,
"Observation of a Laser-Induced Structure in the Ionization
Continuum of Sodium Atoms Using Photoelectron Energy
Spectroscopy," J. Phys. B \textbf{ 30}, 3789-96 (1997)


\bibitem{Kimb} A. K. Popov and V.V. Kimberg, "Nonlinear-Optical Generation of
Short-Wavelength Radiation Controlled by Laser-Induced
Interference Structures,"  Quantum Electron. \textbf{ 28},  228-34
(1998) [translated from Kvantovaya Elektronika \textbf{ 25},
236-42 (1998)]; \url{http://turpion.ioc.ac.ru/}.



\bibitem{Fau99} O. Faucher, E. Hertz, B. Lavorel, R. Chaux, T. Dreier, H. Berger and D.
Charalambidis," Observation of Laser-induced Continuum Structure
in the NO Molecule," J. Phys. B: At. Mol. Opt. Phys. \textbf{ 32},
4485–93 (1999).






\bibitem{D1}  U. Gaubatz, P. Rudecki, S. Schiemann, and K. Bergmann, "Population
Transfer Between Molecular Vibrational Levels by Stimulated Raman
Scattering with Partially  Overlapping Laser Fields. A New Concept
and Experimental Results," J. Chem. Phys. \textbf{ 92}, 5363-76
(1990).

\bibitem{D2}  S. Schiemann, A. Kuhn, S. Steuerwald, and K.
Bergmann, "Coherent Population Transfer in NO with Pulsed Lasers,"
Phys. Rev. Lett. \textbf{ 71}, 3637-40 (1993).


\bibitem{U1}  M. V. Danileyko, V. I. Romanenko, and L. P. Yatsenko,
"Landau-Zener Transition and Population Transfer in the
Three-Level System Driven by Two Delayed Laser Pulses," Opt.
Commun. \textbf{ 109}, 462-6 (1994).

\bibitem{U2} M. V. Danileiko, V. I. Romanenko, and L. P. Yatsenko, "Adiabatic
Population Transfer on Magnetic Sublevels of Molecules - A New
Possibility of its Motion Control," Ukraine Phys. J. \textbf{ 40},
665-9 (1995).

\bibitem{1D} B. W. Shore, J. Martin, M. Fewell, and K. Bergmann, "Coherent Population Transfer
in Multilevel Systems with Magnetic Sublevels. I. Numerical
Studies," Phys. Rev. A \textbf{ 52}, 566-82 (1995).

\bibitem{1DD} J. Martin, B. W. Shore, and K. Bergmann, "Coherent Population Transfer
in Multilevel Systems with Magnetic Sublevels. II. Algebraic
Analysis," Phys. Rev A \textbf{ 52}, 583 (1995).

\bibitem{D4}  J. Martin , B. W. Shore, and K. Bergmann, "Coherent Population Transfer
in Multilevel Systems with Magnetic Sublevels. III. Experimental
Results," Phys. Rev. A \textbf{ 54}, 1556-69 (1996).

\bibitem{D3}  T. Halfmann and K. Bergmann, "Coherent Population Transfer and Dark
Resonances in Sulphur Dioxide Molecules,"  J. Chem. Phys. \textbf{
104}, 7068-82 (1996).


\bibitem{D5}  A. Kuhn, S. Steuerwald, and K. Bergmann, "Coherent Population Transfer
in NO with Pulsed Laser: The Consequences of Hyperfine Structure,
Doppler Broadening and Electromagnetically Induced Absorption,"
Europ. Phys. J. D \textbf{ 1}, 57-70 (1998).

\bibitem{D8} K. Bergmann, H. Theuer, and B. W. Shore, "Coherent Population Transfer
Among Quantum States of Atoms and Molecules," Rev. Mod. Phys.
\textbf{ 70}, 1003-26 (1998).

\bibitem{D9}  N. V. Vitanov, B. W. Shore, and K. Bergmann, "Adiabatic Population
Transfer in Multistate Chains via Dressed Intermediate States,"
Europ. Phys. J. D \textbf{ 4}, 15-29 (1998).

\bibitem{R2} M. N. Kobrak and S. A. Rice, "Equivalence of the Kobrak-Rice
Photoselective Adiabatic Passage and the Brumer-Shapiro Strong
Field Methods for Control of Product Formation in a Reaction," J.
Chem. Phys. \textbf{ 109}, 1-10 (1998).

\bibitem{R3} M. N. Kobrak and S. A. Rice, "Coherent Population Transfer via a Resonant
Intermediate State: The Breakdown of Adiabatic Passage," Phys.
Rev. A \textbf{ 57}, 1158-63  (1998).


\bibitem{102} I. R. Sola, V. S. Malinovsky, B. Y. Chang, J. Santamaria, and
K. Bergmann, "Coherent Population Transfer in Three-Level
Lambda-Systems by Chirped Laser Pulses: How to Minimize the
Population of the Intermediate Level," Phys. Rev. A \textbf{ 59},
4494-4501 (1999).


\bibitem{Band} S. Kallush and Y. B. Band, "Short-pulse Chirped Adiabatic Population
Transfer in Diatomic Molecules," Phys. Rev. A \textbf{ 61},
041401R-1-4 (2000).

\bibitem{Ric00} T. Rickes, L. P. Yatsenko, S.
Steuerwald, T. Halfmann, B. W. Shore, N. V. Vitanov, and K.
Bergmann, "Efficient Adiabatic Population Transfer by Two-Photon
Excitation Assisted by a Laser-Induced Stark shift," J. Chem.
Phys. \textbf{ 113}, 534-546 (2000).

\bibitem{Br6} M. Shapiro and P. Brumer, "On the Origin of Pulse Shaping Control of
Molecular Dynamics,"  J. Chem. Phys. A \textbf{ 105}, 2897-2902,
(2001).

\bibitem{R7} V. Kurkal and S. A. Rice, "Sensitivity of the Extended STIRAP Method of
Selective Population Transfer to Coupling to Background States,"
J. Phys. Chem. B \textbf{ 105}, 6488-94 (2001).

\bibitem{117} N. V. Vitanov, M. Fleischhauer, B. W. Shore, and K. Bergmann, "Coherent
Manipulation of Atoms and Molecules by Sequential Pulses," in
\textit{ Advances of Atomic, Molecular, and Optical Physics}, Vol.
\textbf{ 46}, edited by B. Bederson and H. Walther  (Academic
Press, San Diego, 2001), pp. 55-190.

\bibitem{YatOC02}  L. P. Yatsenko, N. V. Vitanov, B. W. Shore, T. Rickes,
and K. Bergmann, "Creation of Coherent Superposition Using
Stark-Chirped Rapid Adiabatic Passage," Opt. Commun. \textbf{
204}, 413-423 (2002).



\bibitem{Lam94} T. Nakajima, M. Elk, J. Zhang, and P. Lambropolous, "Population
Transfer Through the Continuum," Phys. Rev. A \textbf{ 50},
R913-16 (1994).

\bibitem{Kn97} E. Paspalakis, M. Protopapas, and P. L. Knight, "Population
Transfer Through the Continuum with Temporally Delayed Chirped
Laser Pulses," Opt. Commun. \textbf{ 142}, 34-40 (1997).

\bibitem{U5}  L. Yatsenko, R. Unanyan, K. Bergmann, T. Halfmann, and B. W. Shore,
"Population Transfer through the Continuum Using Laser-Controlled
Stark Shifts," Opt. Commun. \textbf{ 135}, 406-412 (1997).

\bibitem{Kn98} E. Paspalakis, M. Protopapas, and P. L. Knight,
"Time-Dependent Pulse and Frequency Effects in Population Trapping
via the Continuum," J. Phys. B \textbf{ 31}, 775-94 (1998)


\bibitem{A5}  R. G. Unanyan, N. V. Vitanov, and S. Stenholm, "Suppression of Incoherent
Ionization in Population Transfer via Continuum," Phys. Rev. A
\textbf{ 57}, 462-6 (1998).


\bibitem{Half}T. Halfmann, L. P. Yatsenko, M. Shapiro, B. W. Shore, and K. Bergmann,
"Population Trapping and Laser-Induced Continuum Structure in
Helium: Experiment and Theory," Phys. Rev. A \textbf{ 58}, R46-49
(1998) .


\bibitem{Yats} L. P. Yatsenko, T. Halfmann, B. W. Shore, and K. Bergmann,
"Photoionization Suppression by Continuum Coherence: Experiment
and Theory," Phys. Rev. A  \textbf{ 59}, 2926-2947 (1999).


\bibitem{U9}  L. P. Yatsenko, T. Halfmann, B. W. Shore, and K. Bergmann, "Photoionization
Suppression by Continuum Coherence: Experiment and Theory," Phys.
Rev. A \textbf{  59}, 2926-47 (1999).


\bibitem{105} R. G. Unanyan, N. V. Vitanov, B. W. Shore, and K. Bergmann, "Coherent
Properties of a Tripod System Coupled via a Continuum," Phys. Rev.
A \textbf{ 61}, 043408-1-10 (2000).



\bibitem{Sh94} M. Shapiro, "Theory of One-and Two-Photon Dissociation with
Strong Laser Pulses," J. Chem.  Phys. \textbf{ 101}, 3844-51
(1994).

\bibitem{Sh95} Z. D. Chen, M. Shapiro, and P. Brumer, "Incoherent
Interference Control of Two-Photon Dissociation," Phys. Rev. A
\textbf{ 52}, 2225-33 (1995).
\bibitem{Sh96} E. Frishman and M. Shapiro, "Reversibility of Bound-to-Continuum
Transitions Induced by a Strong Short Laser Pulse and the
Semiclassical Uniform Approximation," Phys. Rev. A \textbf{ 54},
3310-21 (1996).


\bibitem{ShV96}A. Vardi and M. Shapiro, "Two-Photon Dissociation/Ionization
Beyond the Adiabatic Approximation," J. Chem. Phys. \textbf{ 104},
5490-6 (1996).

 \bibitem{Sh97} A. Vardi, D. Abrashkevich, E. Frishman, and M. Shapiro, "Theory
of Radiative Recombination with Strong Laser Pulses and the
Formation of Ultracold Molecules via Stimulated
Photo-Recombination of Cold Atoms," J. Chem. Phys. \textbf{ 107},
6166-74 (1997).


\bibitem{Berg} A. Vardi, M. Shapiro, and K. Bergmann, "Complete Population Transfer to
and from a Continuum and the Radiative Association of Cold Na
Atoms to Produce Translationally Cold Na$_2$ Molecules in Specific
Vib-Rotational States," Optics Express \textbf{ 4}, 91-106 (1999).



\bibitem{L22} I. Pastirk, E. J. Brown, I. Grimberg, V. V. Lozovoy, and M. Dantus, "Sequence for
Controlling Laser Excitation with Femtosecond Three-Pulse
Four-Wave Mixing," Faraday Discuss.  \textbf{ 113}, 401-24 (1999).

\bibitem{L23} E. J. Brown, I. Pastirk, V. V. Lozovoy, I. Grimberg,
and M. Dantus, "Population and Coherence Control by Three-Pulse
Four-Wave Mixing,"  J. Chem. Phys. \textbf{ 111}, 3779-82 (1999).

\bibitem{L24} I. Pastirk, V. V. Lozovoy, E. J. Brown, I. Grimberg, and M. Dantus, "Control and
Characterization of Intramolecular Dynamics with Femtosecond
Three-Pulse Four-Wave Mixing," J. Phys. Chem. \textbf{ 103},
10226-36 (1999).

\bibitem{L25} V. V. Lozovoy, B. I. Grimberg, E. J. Brown, I. Pastirk, and M. Dantus, "Femtosecond
Spectrally Dispersed Three-Pulse Four-Wave Mixing: the Role of
Sequence and Chirp in Controlling Intramolecular Dynamics," J.
Raman Spectrosc. \textbf{ 31}, 41-49 (2000).


\bibitem{L28} V. V. Lozovoy, E. J. Brown, I. Pastirk, B. I. Grimberg, and M. Dantus, "What Role
Can Four-Wave Mixing Techniques Play in Coherent Control," in
\textit{ Advances in Multiphoton Processes and Spectroscopy}, Vol.
\textbf{ 14},  edited by R. J. Gordon and Y. Fujimura (World
Scientific, Singapore, 2000), pp. 62-79.

\bibitem{L29} V. V. Lozovoy, I. Pastirk, E. J. Brown, B. I. Grimberg, and M. Dantus, "The Role of
Pulse Sequences in Controlling Ultrafast Intramolecular Dynamics
with Four-Wave-Mixing," Int. Rev. Phys. Chem. \textbf{ 19},
531-52. (2000).

\bibitem{L32} V. V. Lozovoy, I. Pastirk, M. G. Comstock, and M. Dantus, "Cascaded Optical
Free-Induction Decay Four-Wave Mixing," Chem. Phys. Lett. \textbf{
266}, 205-12 (2001).


\end{thebibliography}

\end{document}